\documentclass[journal,letterpaper,twoside,twocolumn]{IEEEtran}
\usepackage{amsmath,amssymb,graphicx,psfrag,cite}
\usepackage[normalem]{ulem} 
\usepackage[usenames,dvipsnames]{color} 
\usepackage{balance}
\usepackage{mathrsfs}   
\usepackage{paralist}

\graphicspath{{figure/}}

\usepackage{color}

\newcommand{\rev}[1]{{\color{blue}#1}}

\definecolor{emphasis}{RGB}{204,0,204}

 \renewcommand{\sout}[1]{} 
  \renewcommand{\rev}[1]{#1}

\title{Compressed Shaping:\\Concept and FPGA Demonstration} %
\author{Tsuyoshi~Yoshida,~\IEEEmembership{Member,~IEEE,} 
	Koji~Igarashi,~\IEEEmembership{Member,~IEEE,} 
	Magnus~Karlsson,~\IEEEmembership{Fellow,~OSA;~Senior~Member,~IEEE,}
	and Erik~Agrell,~\IEEEmembership{Fellow,~IEEE} 
	\thanks{This work was presented in part at OFC 2019\cite{yoshida_2019_ofc} and ECOC 2019\cite{yoshida_2019_ecoc_fpga}.}
	\thanks{T.~Yoshida is with Information Technology R\&D Center, Mitsubishi Electric Corporation, Kamakura, 247-8501, Japan. He also belongs to Graduate School of Engineering, Osaka University, Suita, 565-0871, Japan (e-mail:  Yoshida.Tsuyoshi@ah.MitsubishiElectric.co.jp).}
	\thanks{K.~Igarashi is with Graduate School of Engineering, Osaka University, Suita, 565-0871, Japan.}
	\thanks{M.~Karlsson is with the Dept.~of Microtechnology and Nanoscience and E.~Agrell is with the Dept.~of Electrical Engineering, both at Chalmers University of Technology, SE-41296 Gothenburg, Sweden.}
	\thanks{This work was partly supported by ``Massively Parallel and Sliced Optical Network (MAPLE),'' the Commissioned Research of National Institute of Information and Communications Technology (NICT), Japan (project no.~20401).}
	\thanks{Copyright (c) 2021 IEEE. Personal use of this material is permitted. However, permission to use this material for any other purposes must be obtained from the IEEE by sending a request to pubs-permissions@ieee.org.}
}%

\begin{document}
\maketitle

\begin{abstract}
Probabilistic shaping (PS) has been widely studied and applied to optical fiber communications. The encoder of PS expends the number of bit slots and controls the probability distribution of channel input symbols. Not only studies focused on PS but also most works on optical fiber communications have assumed source uniformity (i.e. equal probability of  marks and spaces) so far. On the other hand, the source information is in general nonuniform, unless bit-scrambling or other source coding techniques to balance the bit probability is performed. Interestingly, one can exploit the source nonuniformity to reduce the entropy of the channel input symbols with the PS encoder, which leads to smaller required signal-to-noise ratio at a given input logic rate. This benefit is equivalent to a combination of data compression and PS, and thus we call this technique \emph{compressed shaping}. In this work, we explain its theoretical background in detail, and verify the concept by both numerical simulation and a field programmable gate array (FPGA) implementation of such a system. In particular, we find that compressed shaping can reduce power consumption in forward error correction decoding by up to 90\% in nonuniform source cases. The additional hardware resources required for compressed shaping are not significant compared with forward error correction coding, and \sout{a real-time back-to-back} \rev{an error insertion} test is successfully demonstrated \rev{with the FPGA.} 
\end{abstract}


\section{Introduction}
\label{sec:intro}

Traffic demands are growing with deployments of mobile communication systems for the 5th generation and beyond\sout{.}\rev{, which is further accelerated by the pandemic \cite{feldmann_2020}.}
Optical fiber communications take a key role in the communication infrastructure because of its high capacity. In the past, the modulation formats used  in optical fiber communications were binary, e.g., on--off keying, binary, or quaternary phase-shift keying without forward error correction (FEC) or with hard-decision FEC \cite{roberts_2009_jlt}. However, the latest 400 Gb/s standards \cite{400zr,openROADM} utilize 16-ary quadrature amplitude modulation (QAM) with soft-decision (SD) FEC under bit-interleaved coded modulation (BICM)\cite{zehavi_1992_tcom,caire_1998_tit,fabregas_2008,bicmbook}. Furthermore, constellation shaping \cite{forney_1989_jsac}, or more specifically, probabilistic shaping (PS) \cite{calderbank_1990_tit,kschischang_1993_tit}, has attracted wide research interest due to its capacity-approaching performance \cite{bocherer_2015_tcom,buchali_2016_jlt,bocherer_2019_jlt,cho_2019_jlt}. Especially reverse concatenation, where the shaping encoding (also known as \emph{distribution matching}, DM)\cite{schulte_2016_tit,gultekin_2018_isit,fehenberger_2019_tcom,yoshida_2019_jlt,schulte_2019_wcl,cho_2020_tcom}, is done outside the FEC encoding \cite{bliss_1981_ibm,fan_1999_globecom,djordjevic_2006_jlt}, made PS deployable in terms of implementation capability.

An optimal encoder, which minimizes the rate loss in the conversion process, can be theoretically achieved in two steps if the block length is large enough; by applying source coding (often called ``data compression'') first and channel coding (i.e., constellation shaping and FEC coding) next. Information-theoretic coding and modulation techniques have realized significant performance improvements in recent years, thus almost closing the gap to the Shannon channel capacity \cite{shannon_1948}. In contrast, coding for dynamically variable source information\rev{\cite{feldmann_2020}} has rarely been investigated for fiber-optic communication systems, which aggregates massive user traffic in frames. 
In the standard \cite{itu-t_g709}, simple bit scrambling (flipping bits by the exclusive OR operation with a pseudorandom bit sequence (PRBS)) has been implemented to balance the mark (logic `$1$') and space (logic `$0$') counts instead of applying any source coding. Often we tend to assume the source bits as just uniformly distributed and independent, although the true source entropy before bit scrambling is variable and dependent on the user traffic, e.g., due to the existence of idle frames in the media access control \rev{(MAC)} protocol \cite{itu-t_g709}.

Data compression and shaping are almost inverse operations, i.e., the former converts a nonuniform information sequence into a shorter uniform one, while the latter does the opposite. Simultaneous realization of data compression and shaping is not only an interesting research topic but also a key technique for more efficient communications in practice.
Thus in this paper, we propose and investigate \emph{compressed shaping}, which combines the benefits from data compression and shaping.
Similar ideas have been studied in the context of joint source--channel coding in communication theory \cite{sayood_1991_tcom,kliewer_2003_elet}, but an application to fiber-optic communication is presented here, for the first time to the best of our knowledge.

Compressed shaping is enabled by a shaping encoding that is sensitive to the source entropy. As data compression allocates short bit patterns to frequently occurring source words, compressed shaping allocates amplitudes with small energy to such frequent source words. 
This compression feature is similar to burst signalling in time-domain multiple access \cite{vacondio_2013_ecoc}, i.e., optical power variation depending on the traffic.
Compressed shaping is a fixed-length to fixed-length conversion, and the average energy of channel input symbols is reduced for source information sequences having a small entropy.
We do not need an operational mode change such as updating the source statistics based on prior knowledge, although this would be a kind of data compression.
Our previously proposed look-up table (LUT)-based hierarchical DM \cite{yoshida_2019_jlt} and following works \cite{civelli_2020_ofc,civelli_2020_entropy} \sout{are applicable for this purpose without significantly increased complexity, but rather a reordering of the LUT entries.}
\rev{can be applied for this purpose by reordering the LUT entries without significantly increased complexity.}
The proposed technique can reduce the rate losses associated with source and channel coding compared with state-of-the-art DM schemes such as constant-composition DM (CCDM) \cite{schulte_2016_tit}, or reduce the power consumption in the FEC at a given information rate and signal-to-noise ratio (SNR) by relaxing the FEC performance.
We also report a field programmable gate array (FPGA) implementation, \sout{realtime evaluation} \rev{and the hardware demonstration} 
results for compressed shaping $16$- and $64$-QAM at system throughputs of $153$ and $113\,\mathrm{Gb/s}$, respectively.

This is an evolutional work of \cite{yoshida_2019_ofc,yoshida_2019_ecoc_fpga,yoshida_2020_eic}, which are here extended by providing a more detailed theoretical background of compressed shaping. 
The system throughput is increased by separating the clock domain into data and controlling because the control circuitry was the bottleneck in the logical circuitry when making the clock frequency faster.
The accuracy of the power consumption estimation is improved by introducing a dynamic simulation.
Even if we consider shaping encoding only, there have been very few other reports on FPGA implementations \cite{yu_2020_ofc,yu_2020_jlt}. Furthermore, there are neither any reports on static/dynamic power consumption estimates nor any \sout{real-time evaluations of} \rev{demonstrations by} FPGA implementations including both shaping encoding and decoding, except \cite{yoshida_2019_ecoc_fpga}.

The rest of the paper is organized as follows. The principle of compressed shaping is explained in Sec.~\ref{sec:principle}, and its numerical simulations are shown in Sec.~\ref{sec:sim}.
An FPGA implementation example of compressed shaping is presented in Sec.~\ref{sec:fpga}, and \sout{real-time} 
\rev{the hardware-based demonstration} results are summarized in Sec.~\ref{sec:demo}. Finally, Sec.~\ref{sec:sum} concludes the paper.

\section{Compressed shaping---basic principles}
\label{sec:principle}
While conventional PS systems employ full bit-scrambling and assume source uniformity, the proposed compressed shaping scrambles the sign bit only and applies source-sensitive amplitude shaping to realize better performance in the case of small source entropy.
In this section, we firstly review the historical source uniformity assumption in conventional systems and discuss rate loss under either uniform or nonuniform source conditions. Then we compare compressed shaping with bit-sequence data compression and PS. Finally, we show the system model and characterize entropy bounds in compressed shaping systems.

\subsection{Rate loss in conventional systems}
\rev{Optical fiber communication systems gather many} \sout{media access control} \rev{MAC} \rev{frames from client traffic.}
When investigating coding and modulation techniques, source uniformity is usually assumed. 
It is because a $50\%$ mark ratio is produced by \sout{transcoding} \rev{line coding} or bit scrambling, even if the true source information is nonuniform.
\rev{This may be a reasonable model under the idealized assumption of a stable, uniform traffic flow. However, due to dynamically variable traffic volumes \cite{feldmann_2020},
there will be many idle frames.}
\sout{There are usually idle frames, which}
These are \sout{transcoded} \rev{encoded} by 64B/66B line coding into the zero codeword except for \sout{the control bits} \rev{the sync header and block type field} \cite{itu-t_g709}. Due to the existence of such idle frames, the source bits $S_i\in\{0,\,1\}$ can have more than $50\%$ `$0$'s, which makes the binary source entropy $\mathbb{H}(S_i) < 1$, where $\mathbb{H}(\cdot)$ denotes entropy.
\rev{For example, when an 8-byte idle signal is encoded by 64B/66B, a 66-bit block consists of the 2-bit sync header $01$, the 8-bit block type field $00011110$, and eight 7-bit control bits $0000000$ for the 8-byte idle signal, which results in a source mark ratio of $5/66=7.6\%$.}

In conventional systems, the (serial) source bits $S_i$ are parallelized to form a source bit sequence $[S_1 \ldots S_{k_{\mathrm{bs}}}]$, after which the bit scrambling converts $[S_1 \ldots S_{k_{\mathrm{bs}}}]$ into a scrambled bit sequence $[U_1 \ldots U_{k_{\mathrm{bs}}}]$. The sequence length $k_{\mathrm{bs}}$ is chosen to match the applied bit-scrambling protocol. An arbitrary (randomly selected) bit in this scrambled sequence has the distribution $P_U = (1/k_{\mathrm{bs}}) \sum_{i=1}^{k_{\mathrm{bs}}} P_{U_i}$, which is assumed to be uniform and $\mathbb{H}(U) = 1$, even when the corresponding (serial) unscrambled bit distribution $P_S = (1/k_{\mathrm{bs}}) \sum_{i=1}^{k_{\mathrm{bs}}} P_{S_i}$ yields an entropy $\mathbb{H}(S) < 1$.
The bit scrambling is essential in binary modulation for maintaining direct current levels in electronic devices and for recovering the clock signal at the receiver. Non-PS QAM systems also utilize bit scrambling to maintain the $50\%$ mark ratio.
In general, neither $S_i$ nor $U_i$ ($i=1,2,\ldots,k_{\mathrm{bs}}$) is ensured to be identical and independent distributed (i.i.d.), so $\mathbb{H}([S_1 \ldots S_{k_{\mathrm{bs}}}]) \le k_{\mathrm{bs}} \mathbb{H}(S)$ and $\mathbb{H}([U_1 \ldots U_{k_{\mathrm{bs}}}]) \le k_{\mathrm{bs}} \mathbb{H}(U)$ by the concavity of entropy. However, $[U_1 \ldots U_{k_{\mathrm{bs}}}]$ is assumed to be i.i.d.\ and uniform in most research works on channel coding.

In PS, the goal is to reduce symbol entropy at a given information rate in order to reduce the required SNR for quasi-error-free operation over channels approximated by the Gaussian channel.
The scrambled bit sequences $[U_1 \ldots U_{k_{\mathrm{bs}}}]$ are rearranged into sequences of length $k$, and each such length-$k$ bit sequence is mapped into a sequence of amplitudes $[A_1 \ldots A_n]$. The sequence lengths $k$ and $n$ are selected to match the PS scheme, regardless of $k_{\mathrm{bs}}$. Since $U_i$ is in general not i.i.d., neither is $A_i$, and thus $\mathbb{H}([A_1 \ldots A_n]) \le n \mathbb{H}(A)$, where $P_A = (1/n) \sum_{i=1}^n P_{A_i}$. On the other hand, $A_i$ is assumed to be i.i.d.\ in a mismatched (memoryless) receiver.

The PS performance is typically quantified with a rate loss
\begin{IEEEeqnarray}{rCL}
	\label{eq:rloss_ps}
	R_{\mathrm{loss}} (U,A) &=& \mathbb{H}(A) - \frac{k}{n},
\end{IEEEeqnarray}
where $R_{\mathrm{loss}} (\cdot,\cdot)$ denotes a rate loss in a sequence conversion, and $R_{\mathrm{loss}}(U,A) \ge 0$ for i.i.d.\ $U_i$.
The CCDM \cite{schulte_2016_tit}, which has been the state-of-the-art PS coding, has an almost negligible rate loss when $n$ is sufficiently large, e.g., $1000$--$10000$.
The CCDM generates a fixed probability mass function (PMF) $P_{A}$ of output amplitudes $A$ regardless of the DM encoder input bit sequence and its statistics.
For a fixed PMF $P_{A}$ with a nonconstant-composition DM, $U_i$ is required to be uniformly distributed in general, which can be achieved by bit scrambling.

\subsection{Rate loss in source to amplitude conversion}
\label{subsec:rloss_gen}
The CCDM shows negligible rate loss $R_{\mathrm{loss}} (U,A)$ in \eqref{eq:rloss_ps} with a sufficiently large block length in PS for a uniform source, but not for nonuniform sources, in which case the channel input symbol entropy or average symbol energy can be further reduced by exploiting nonconstant-composition DM.

During a conversion process from source bit sequence to amplitude symbol sequence, bit scrambling acts against the symbol entropy reduction because it maximizes binary entropy ($\mathbb{H}(U)=1$) even if $\mathbb{H}(S)<1$. Instead, we propose to exploit this source nonuniformity.
In the case of $\mathbb{H}(S) < 1$, the rate loss in the conversion from a nonuniform source bit sequence $[S_1 \ldots S_k]$ into an amplitude sequence $[A_1 \ldots A_n]$ 
can in general be expressed as
\begin{IEEEeqnarray}{rCL}
	\label{eq:rloss_cc}
	R_{\mathrm{loss}} ([S_1 \ldots S_k], A) &=& \mathbb{H}(A) - \frac{ \mathbb{H}([S_1 \ldots S_k])}{n} ,
\end{IEEEeqnarray}
which is bounded as 
\begin{IEEEeqnarray}{rCL}
	R_{\mathrm{loss}} ([S_1 \ldots S_k], A) & \ge & R_{\mathrm{loss}} (S, A) \\
	\label{eq:rloss_cc2}
	R_{\mathrm{loss}} (S,A) & = & \mathbb{H}(A) - \mathbb{H}(S)\frac{k}{n} .
\end{IEEEeqnarray}
Obviously, CCDM is not optimum (with or without bit scrambling) because of the constant $P_{A}$ and $\mathbb{H}(A)$ regardless of the source distribution. The general rate loss bound $R_{\mathrm{loss}} (S, A)$ in \eqref{eq:rloss_cc2} becomes significantly larger than $R_{\mathrm{loss}} (U, A)$ in \eqref{eq:rloss_ps} under a small $\mathbb{H}(S)$. 
A nonconstant-composition DM could be source sensitive, i.e., realizing a smaller output entropy $\mathbb{H}(A)$ for a smaller input entropy $\mathbb{H}(S)$, resulting in a smaller $R_{\mathrm{loss}} (S,A)$ compared with CCDM. Even in such cases, at least the sign bits should be bit-scrambled for the direct current level management and clock recovery.
To enhance the performance under nonuniform source information, data compression is another option, at the expense of, possibly large, digital signal processing circuit resources, since it has to adapt to the time variation of $\mathbb{H}(S)$.

\subsection{Proposed bit sequence conversion}
This section explains the principle of the proposed compressed shaping compared to well-known data compression and PS. Tab.~\ref{tab:cc} shows small examples of bit sequence conversions using (a) data compression, (b) PS, and (c) compressed shaping.

Tab.~\ref{tab:cc}(a) assumes a nonuniform source input bit sequence $\boldsymbol{S}=[S_1 S_2]$ with a fixed length $k=2$. The bit sequence conversion is given by Huffman coding \cite{huffman_1952}, which allocates a short output string to a high probability input word. The output bit sequence $\boldsymbol{C}$ have variable lengths $n_{\boldsymbol{C}}$ from $1$ to $3$ in this example. Then the fixed input length $k=2$ is shortened to $n_{\mathrm{avg}}=\mathbb{E}[n_{\boldsymbol{C}}]=1.7$ on average, where $\mathbb{E}$ denotes expectation. Such conversion is useful to reduce the required storage size after conversion. Huffman coding is an invertible data compression; however, it is usually not suitable for high throughput data communications due to issues of latency and required storage size in the variable-length conversion process.

Tab.~\ref{tab:cc}(b) assumes a uniform source input bit sequence $\boldsymbol{U}=[U_1 U_2]$ with a fixed length $k=2$, which is converted into an amplitude sequence $\boldsymbol{A}=[A_1 A_2 A_3]$ with a fixed length $n=3$ and an amplitude element $A \in \{1,\,3\}$.  
Candidates of output amplitudes are sorted \sout{by} \rev{in} ascending order of average energy per codeword $E_{\boldsymbol{A}}=|| \boldsymbol{A} ||^2_2 / n$.
The output amplitude sequence $\boldsymbol{A}=111$ has the smallest $E_{\boldsymbol{A}}$ of $1$, and $\boldsymbol{A}=113$, $131$, and $311$ have the second smallest $E_{\boldsymbol{A}}$ of $11/3$. 
Amplitude sequences $\boldsymbol{A}=133$, $313$, $331$, and $333$ are not chosen as output strings in this codebook because of their large $E_{\boldsymbol{A}}$. Then the average output energy $E_{\mathrm{avg}}=\mathbb{E}[E_{\boldsymbol{A}}] = 3$.
When we use a codebook with uniform output amplitudes of $1$ and $3$ with a length of $2$ (i.e., $11$, $13$, $31$, and $33$), $E_{11}=1$, $E_{13}=E_{31}=5$, $E_{33}=9$, and $E_{\mathrm{avg}}=5$. Compared with such a uniform output amplitude case, $E_{\mathrm{avg}}$ is reduced by $2.2\,\mathrm{dB}$ with this exemplified PS, leading to a small required SNR at a given information rate over the Gaussian channel.
To shape the output amplitude probabilities (i.e., to reduce the output amplitude entropies), we thus need more output bit slots than input bit slots.

Tab.~\ref{tab:cc}(c) exemplifies the proposed compressed shaping, which assumes a nonuniform source input bit sequence $\boldsymbol{S}=[S_1 S_2]$ as in the case of Tab.~\ref{tab:cc}(a). The source bit sequence $\boldsymbol{S}$ with a fixed length $k=2$ is converted into an amplitude sequence $\boldsymbol{A}=[A_1 A_2 A_3]$ with a fixed length $n=3$. When generating the codebook, we need two sortings: (i) the input bit sequences $\boldsymbol{S}$ are sorted \sout{by} \rev{in} descending order of probability $P_{\boldsymbol{S}}$, and (ii) the output amplitude sequence $\boldsymbol{A}$ is sorted \sout{by} \rev{in} ascending order of average energy $E_{\boldsymbol{A}}$. By allocating output amplitude sequences with a small energy to input bit sequences with a high probability, the average output energy $E_{\mathrm{avg}}$ becomes $7/3$, which is $1.1\,\mathrm{dB}$ further less than the one in Tab.~\ref{tab:cc}(b). 
Both conventional PS and compressed shaping are fixed-length to fixed-length conversions in these examples, and are therefore suitable for high-throughput data communication which arranges data into fixed-length frames. We do not need any adaptation of the codebook to different source statistics.

\begin{table}[t]
	\caption{Bit sequence conversions.}
	\label{tab:cc}
	\vspace{-0.4cm}
	\begin{center}
	(a) Data compression \\
	\begin{tabular}{|c|c|c|c|c|}
		\hline
		\multicolumn{2}{|c|}{Input} & \multicolumn{3}{|c|}{Output} \\ \hline
		Bits & Probability & Bits & Probability & Length \\
		$\boldsymbol{S}$ & $P_{\boldsymbol{S}}$ & $\boldsymbol{C}$ & $P_{\boldsymbol{C}}$ & $n_{\boldsymbol{C}}$ \\
		\hline\hline
		00 & 0.50 & 1 & 0.50 & 1 \\
		01 & 0.30 & 01 & 0.30 & 2 \\
		10 & 0.15 & 001 & 0.15 & 3 \\
		11 & 0.05 & 000 & 0.05 & 3 \\ \hline
		\multicolumn{4}{|c|}{Average} & 1.7 \\ \hline
	\end{tabular}
	\vspace{0.2cm}
	\\ (b) PS \\
	\begin{tabular}{|c|c|c|c|c|}
		\hline
		\multicolumn{2}{|c|}{Input} & \multicolumn{3}{|c|}{Output} \\ \hline
		Bits & Probability & Amplitudes & Probability & Avg. energy \\
	$\boldsymbol{U}$ & $P_{\boldsymbol{U}}$ & $\boldsymbol{A}$ & $P_{\boldsymbol{A}}$ & $E_{\boldsymbol{A}}$ \\
		\hline\hline
		00 & 0.25 & 111 & 0.25 & 1 \\
		01 & 0.25 & 113 & 0.25 & 3.67 \\
		10 & 0.25 & 131 & 0.25 & 3.67 \\
		11 & 0.25 & 311 & 0.25 & 3.67 \\ \hline
		\multicolumn{4}{|c|}{Average} & 3 \\ \hline
	\end{tabular}
	\vspace{0.2cm}
	\\ (c) Compressed shaping \\
	\begin{tabular}{|c|c|c|c|c|}
		\hline
		\multicolumn{2}{|c|}{Input} & \multicolumn{3}{|c|}{Output} \\ \hline
		Bits & Probability & Amplitudes & Probability & Avg. energy \\
		$\boldsymbol{S}$ & $P_{\boldsymbol{S}}$ & $\boldsymbol{A}$ & $P_{\boldsymbol{A}}$ & $E_{\boldsymbol{A}}$ \\
		\hline\hline
		00 & 0.50 & 111 & 0.50 & 1 \\
		01 & 0.30 & 113 & 0.30 & 3.67 \\
		10 & 0.15 & 131 & 0.15 & 3.67 \\
		11 & 0.05 & 311 & 0.05 & 3.67 \\ \hline 
		\multicolumn{4}{|c|}{Average} & 2.33 \\ \hline
	\end{tabular}
	\end{center}
\end{table}

\subsection{Proposed system model}
\label{subsec:sys}
\begin{figure}[t]
	\begin{center}
		\setlength{\unitlength}{.6mm} %
		\scriptsize
		\vspace{-0.1cm}
		\includegraphics[scale=0.34]{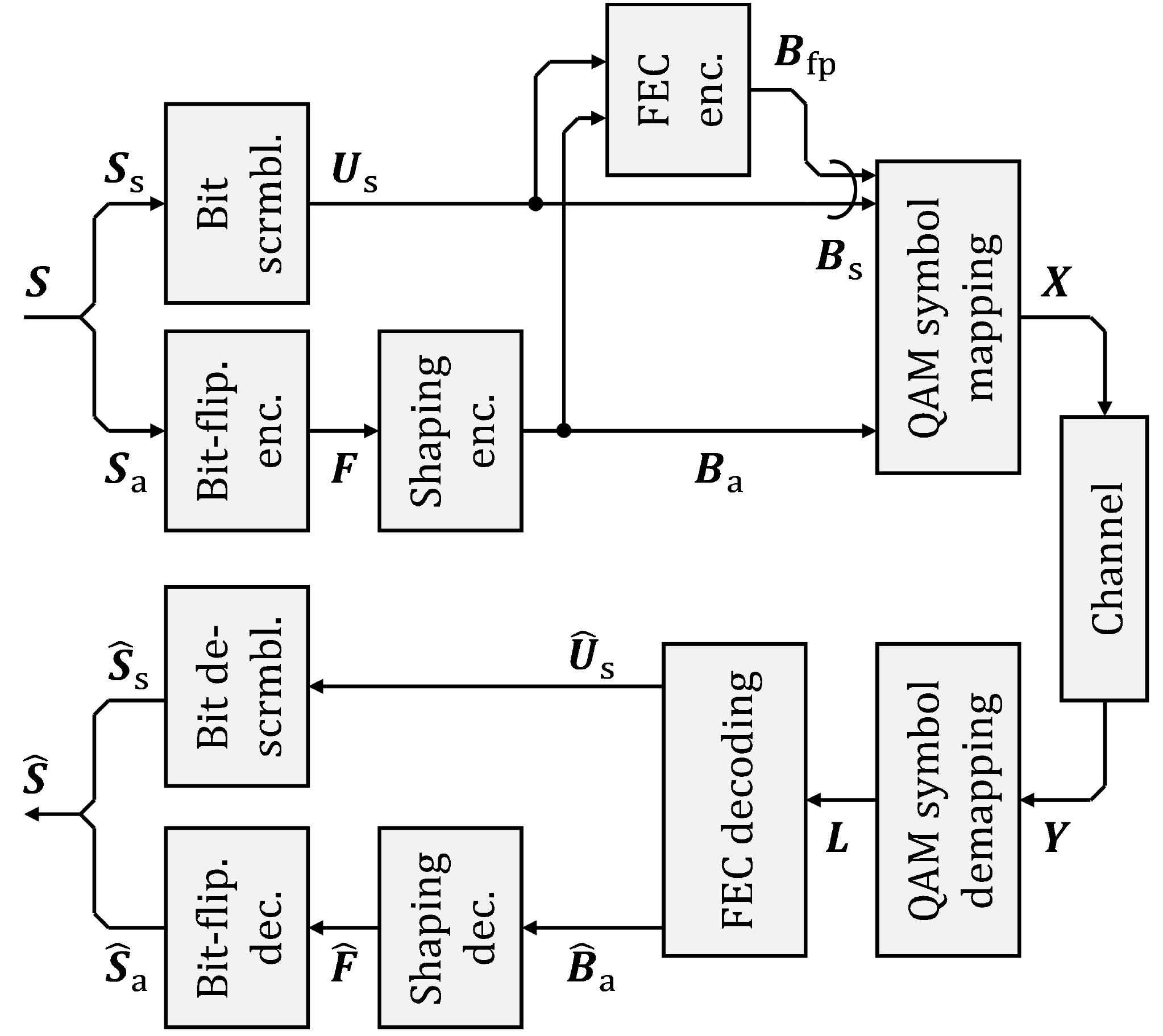} \\
		\vspace{-0.3cm}		
		\caption{System model for the proposed compressed shaping.}
		\label{fig:sys}
	\end{center}
	\vspace{-0.4cm}
\end{figure}

\begin{figure}[t]
	\begin{center}
		\setlength{\unitlength}{.6mm} %
		\scriptsize
		\vspace{-0.1cm}
		\includegraphics[scale=0.36]{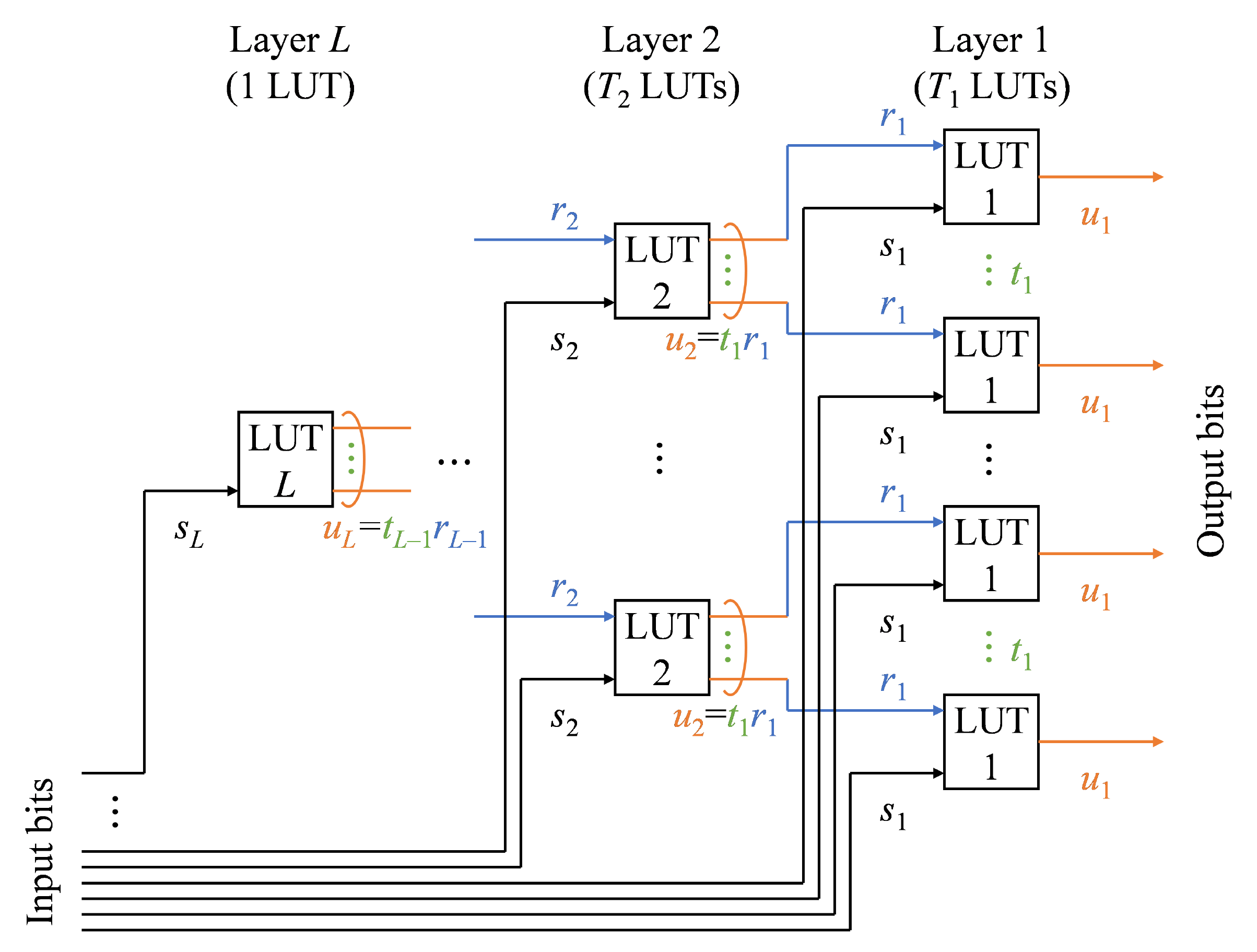} \\
		\vspace{-0.3cm}
		\caption{Schematic of hierarchical DM encoding \cite[Fig.~2]{yoshida_2020_izs}.}
		\label{fig:hidm}
	\end{center}
	\vspace{-0.4cm}
\end{figure}

Fig.~\ref{fig:sys} shows the system model for the proposed compressed shaping. Here we exemplify using hierarchical DM \cite{yoshida_2019_jlt} for the shaping encoder/decoder, but generally any nonconstant-composition DM can be used.
The source information bits are parallelized into a source bit sequence $\boldsymbol{S}=[S_1 \ldots S_{k_{\mathrm{bs}}+k}]$, which is separated into a source sign bit sequence $\boldsymbol{S}_{\mathrm{s}}=[S_{\mathrm{s},1} \ldots S_{\mathrm{s},k_{\mathrm{bs}}}]$ and a source amplitude bit sequence $\boldsymbol{S}_{\mathrm{a}}=[S_{\mathrm{a},1} \ldots S_{\mathrm{a},k}]$.
The source sign bit sequence $\boldsymbol{S}_{\mathrm{s}}$ is bit-scrambled into $\boldsymbol{U}_{\mathrm{s}}=[U_{\mathrm{s},1} \ldots U_{\mathrm{s},k_{\mathrm{bs}}}]$ by taking the exclusive OR with a PRBS, to balance the numbers of `$0$'s and `$1$'s. The source amplitude bit sequence $\boldsymbol{S}_{\mathrm{a}}$ is processed by a bit-flipping function, which flips all input bits if there are more `$1$'s than `$0$'s and adds a parity bit `$1$', otherwise just adds a parity bit `$0$', because a large mark ratio is not desirable in the compressed shaping scheme.\footnote{There may be situations with a predominance of `$0$'s (due to many idle frames) or `$1$'s (due to an alarm indication signal) \cite{itu-t_g709}. In both cases, the source entropy is small.}
Then the bit-flipping encoding output bit sequence $\boldsymbol{F}=[F_1 \ldots F_{k+1}]$ contains at least as many `$0$'s as `$1$'s.
The sequence $\boldsymbol{F}$ is then processed in a hierarchical DM, which consists of hierarchically connected small look-up tables (LUTs) as shown in Fig.~\ref{fig:hidm} \cite{yoshida_2019_jlt,yoshida_2020_izs}. 
There are $L$ layers and $T_{\ell}$ LUTs in each layer $\ell$. Each LUT in a layer $\ell$  receives $s_{\ell}$ bits from the input interface of the DM and $r_{\ell}$ bits from layer $\ell + 1$, and it transmits $r_{\ell-1}$ bits to each of $t_{\ell -1}$ LUTs in layer $\ell -1$, in total $u_{\ell}=t_{\ell -1} r_{\ell -1}$ transmitted bits.  
To determine the one-to-one correspondence of input and output words in each small LUT, we sort the input words in descending order of the number of `$0$'s and the output amplitudes in ascending order of average energy, as in the small example shown in Tab.~\ref{tab:cc}(c).  
The output amplitude sequence from the hierarchical DM is $\boldsymbol{A}=[A_{\mathrm{c},1} \ldots A_{\mathrm{c},n}]$, where each element is a two-dimensional vector $A_{\mathrm{c},i} \in \{ 1,\, 3,\, \ldots ,\, 2^{m_{\mathrm{a}}/2-1} \}^2$, $m_{\mathrm{a}}$ is the number of bit tributaries for a two-dimensional amplitude, and $P_{A_{\mathrm{c}}}=(1/n)\sum_{i=1}^n P_{A_{\mathrm{c},i}}$. The amplitude sequence $\boldsymbol{A}$ is represented by a bit sequence $\boldsymbol{B}_{\mathrm{a}}=[B_{\mathrm{a},1} \ldots B_{\mathrm{a},n m_{\mathrm{a}}}]$.
From $\boldsymbol{B}_{\mathrm{a}}$ and $\boldsymbol{U}_{\mathrm{s}}$, an FEC parity bit sequence $\boldsymbol{B}_{\mathrm{fp}}$ is generated by a systematic FEC encoder with an FEC code rate $R_{\mathrm{c}}$. Then $\boldsymbol{U}_{\mathrm{s}}$ and $\boldsymbol{B}_{\mathrm{fp}}$ are concatenated into a sign bit sequence $\boldsymbol{B}_{\mathrm{s}} = [B_{\mathrm{s},1} \ldots B_{\mathrm{s},n m_{\mathrm{s}}}]$, where $m_{\mathrm{s}}$ denotes the number of sign-bit tributaries. 
The number of elements in $\boldsymbol{S}_{\mathrm{s}}$ or $\boldsymbol{U}_{\mathrm{s}}$ is $k_{\mathrm{bs}}= n(m_{\mathrm{s}}R_{\mathrm{c}}-m_{\mathrm{a}}(1-R_{\mathrm{c}}))$ and that in $\boldsymbol{B}_{\mathrm{fp}}$ is $(1-R_{\mathrm{c}})n(m_{\mathrm{s}}+m_{\mathrm{a}})$. 
Finally channel input QAM symbols $\boldsymbol{X}=[X_{\mathrm{c},1} \ldots X_{\mathrm{c},n}]$ are generated from $\boldsymbol{B}_{\mathrm{s}}$ and $\boldsymbol{B}_{\mathrm{a}}$, where $P_{X_{\mathrm{c}}}=(1/n)\sum_{i=1}^n P_{X_{\mathrm{c},i}}$.

The channel output QAM symbols $\boldsymbol{Y}=[Y_{\mathrm{c},1} \ldots Y_{\mathrm{c},n}]$ are demapped into an L-value sequence $\boldsymbol{L}= [L_1 \ldots L_{(m_{\mathrm{s}}+m_{\mathrm{a}})n}]$ by memoryless bit-metric decoding \cite{bocherer_2015_tcom},
where $Y_{\mathrm{c},i}\in\mathbb{R}^2$ and $\mathbb{R}$ denotes the real number set.
The receiver-side processing reverses the one on the transmitter side. QAM symbol demapping and FEC decoding are performed to recover the scrambled source sign bit sequence $\hat{\boldsymbol{U}}_{\mathrm{s}}$ and the shaped amplitude bit sequence $\hat{\boldsymbol{B}}_{\mathrm{a}}$. The recovered scrambled source sign bit sequence $\hat{\boldsymbol{U}}_{\mathrm{s}}$ is bit-descrambled into the source sign bit sequence $\hat{\boldsymbol{S}}_{\mathrm{s}}$. The shaping decoder, which is a hierarchical DM decoder here, converts $\hat{\boldsymbol{B}}_{\mathrm{a}}$ into the flipped bit sequence $\hat{\boldsymbol{F}}$. The bit-flipping is terminated based on the parity bit, i.e., if the parity bit is `$1$', all input bits are flipped and the parity bit is removed, otherwise the parity bit is just removed, to obtain the source amplitude bit sequence $\hat{\boldsymbol{S}}_{\mathrm{a}}$. Finally, the source sign bit sequence $\hat{\boldsymbol{S}}_{\mathrm{s}}$ and the source amplitude bit sequence $\hat{\boldsymbol{S}}_{\mathrm{a}}$ are concatenated into the source information bit sequence $\hat{\boldsymbol{S}}$. 

We here summarize the entropy and rate loss in this compressed shaping system. The entropy of the channel input symbol is given by
\begin{IEEEeqnarray}{rCL}
	\label{eq:ent_channel}
	\mathbb{H}(X_{\mathrm{c}}) & = & m_{\mathrm{s}} + \mathbb{H}(A_{\mathrm{c}}) .
\end{IEEEeqnarray}
If $S_1,\ldots,S_{k_{\mathrm{bs}}+k}$ are i.i.d.\ with a distribution $P_S=P_{S_{\mathrm{a}}}=\frac{1}{k_{\mathrm{bs}}+k} \sum_{i=1}^{k_{\mathrm{bs}}+k}P_{S_i}$, then the rate loss in the sequence conversion from $S_{\mathrm{a}}$ to $A_{\mathrm{c}}$ is, similarly to \eqref{eq:rloss_cc2},
\begin{IEEEeqnarray}{rCL}
	\label{eq:rloss_cps}
	R_{\mathrm{loss}}(S_{\mathrm{a}},A_{\mathrm{c}}) &=& \mathbb{H}(A_{\mathrm{c}}) - \mathbb{H}(S) \frac{k}{n} .
\end{IEEEeqnarray}
In order to characterize the performance in Sec.~\ref{sec:sim}, we here define the minimum entropy of a channel input symbol as
\begin{IEEEeqnarray}{rCL}
	\label{eq:min_ent_channel}
	\mathbb{H}_{\mathrm{LB}} & = & m_{\mathrm{s}} + \mathbb{H}(S) \frac{k}{n} .
\end{IEEEeqnarray}

\sout{The time-variable PMF $P_{A_{\mathrm{c}}}$ can cause practical issues with respect to electrical amplitude, optical power, and SNR control in fiber-optic communication systems, although the variation of $P_{A_{\mathrm{c}}}$ per wavelength channel can be statistically relaxed by multiplexing many channels. The analysis of such issues and development of appropriate control methods are deferred to future work.}

\rev{Note that the essential function of compressed shaping is the shaping encoder/decoder based on a nonconstant-composition DM. A small example of this is shown in Tab.~\ref{tab:cc}(c).
Once the shaping encoder/decoder functions are determined, they will not be adapted even if the source statistics would vary dynamically.
This feature helps to ensure a low-complexity compressed shaping.
The other additional functions in Fig.~\ref{fig:sys} are optional. For an implementation with the standardized framing using 64B/66B encoding and the alarm indication signal, the bit-flip encoding/decoding is required. To support bipolar signaling such as QAM for coherent systems, the bit scrambling/descrambling of the sign bit is required to maintain the DC level at zero.
With some adjustments, we can apply compressed shaping to  400ZR\cite{400zr} and openROADM\cite{openROADM}. These employ 256B/257B transcoding after the 64B/66B encoding. In the transcoding, most bits are transparently output, while several bits are shortened or changed. Thus the source nonuniformity from the 66-bit blocks are retained. Before the symbol mapping, the sign and amplitude bits must be properly distributed, and the bit scrambling should be modified to avoid breaking the source nonuniformity. These requirements are reasonably minor in a practical implementation.}

\section{Simulations}
\label{sec:sim}

To verify the concept of compressed shaping, we performed numerical simulations based on the system model in Sec.~\ref{subsec:sys}. 
In this section, nonuniform, independent source information bits $S$ were generated for simplicity, by independent, uniformly distributed pseudorandom numbers from the Mersenne twister. 
For a given target source mark ratio $P_S(1)$, the uniformly distributed pseudorandom numbers ranging from $0$ to $1$ were binarized with a threshold level $1-P_S(1)$, i.e., generating a logic `$0$' for a random number $0$ to $1-P_S(1)$ and a logic `$1$' otherwise. In this simulation, we set a static source mark ratio $P_S(1)$ in a simulation batch for a short period.

\begin{figure}[t]
	\begin{center}
		\setlength{\unitlength}{.6mm} %
		\scriptsize
		\vspace{-0.1cm}
		\includegraphics[scale=0.35]{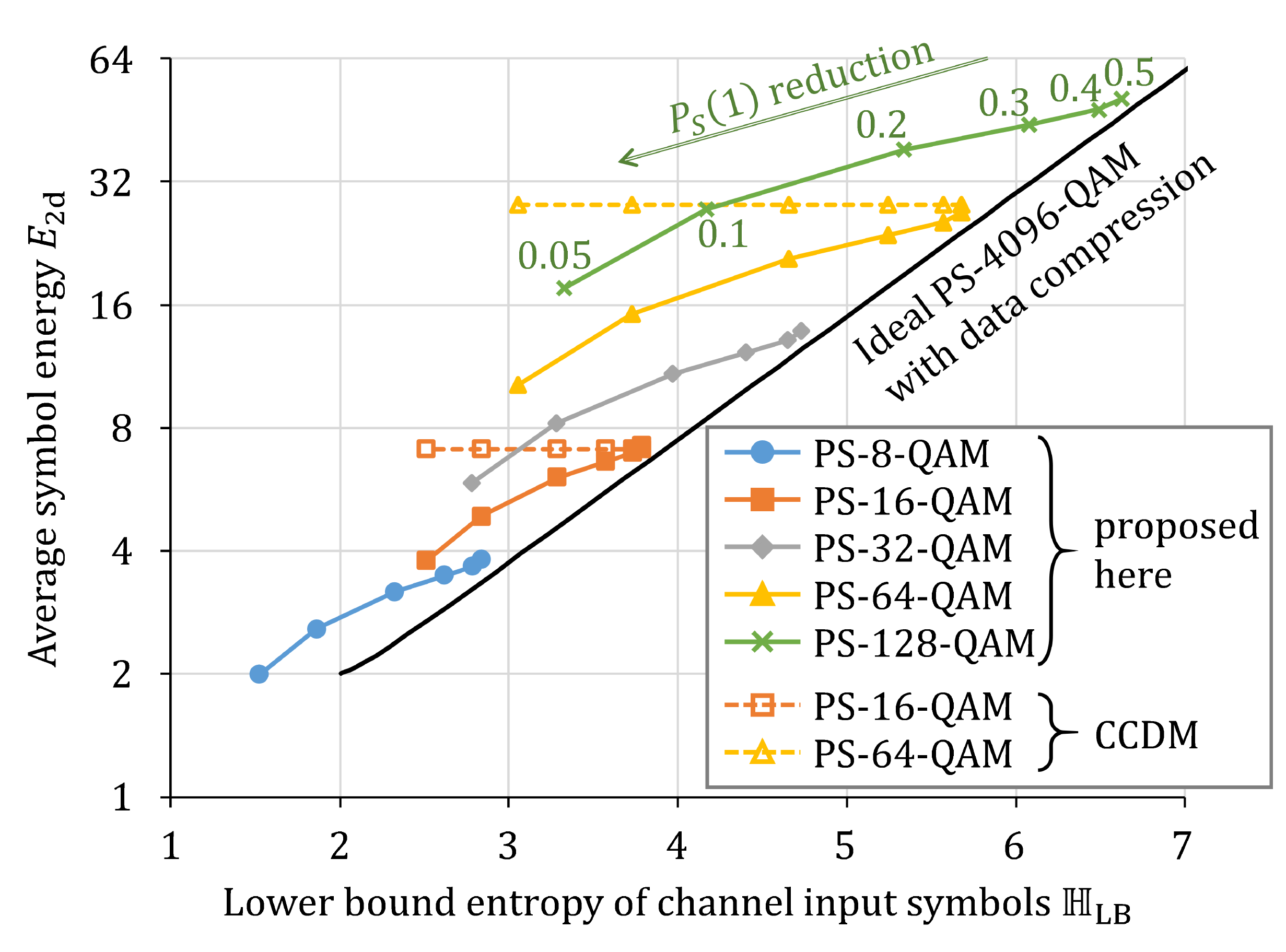} \\
		\vspace{-0.3cm}		
		\caption{Average two-dimensional symbol energy $E_{\mathrm{2d}}$ for various PS-QAM formats and source mark ratios $P_S(1)$ in the range $0.5$--$0.05$ as a function of the lower bound entropy of channel input symbols $\mathbb{H}_{\mathrm{LB}} = m_{\mathrm{s}}+\mathbb{H}(S) k/n$.}
		\label{fig:eavg}
	\end{center}
	\vspace{-0.4cm}
\end{figure}

Fig.~\ref{fig:eavg} shows the average two-dimensional symbol energy $E_{\mathrm{2d}}$ as a function of the lower bound entropy of channel input symbols $\mathbb{H}_{\mathrm{LB}}$ in \eqref{eq:min_ent_channel} for various PS-QAM formats and source mark ratios $P_S(1)=0.5$, $0.4$, $0.3$, $0.2$, $0.1$, and $0.05$, where the minimum Euclidean distance $d_{\mathrm{LB}}=2$.
For compressed shaping with hierarchical DM, we employed $8$-, $16$-, $32$-, $64$-, and $128$-QAM as base constellations. The PS overhead was around $7\%$ in each case when assuming the use of a rate-$5/6$ FEC, and the PS codeword length (number of QAM symbols) was $256$, $256$, $192$, $128$, and $128$ for $8$-, $16$-, $32$-, $64$-, and $128$-QAM, respectively. 
\rev{The PS codeword length equals the block length of compressed shaping. The PS encoder input block consists of several $100$ information bits, and the source mark ratio is averaged over each block. This number of several $100$ bits is small enough to avoid over-averaging of dynamic variations of the source mark ratio due to idle MAC frames.}
Such granular base constellation and shallow shaping help to avoid excessive increases of peak-to-average power ratio and power consumption \cite{zhang_2018_ofc,yoshida_2020_izs} and penalties from nonideal FEC performance \cite{cho_2018_ecoc}.
As the $8$-QAM constellation, $\mathcal{C}_3$ in \cite{schmalen_2017_jlt} was used to make the constellation symmetric around the imaginary axis,\footnote{The signal points are $(\pm 1,\, 0)$, $(\pm 1,\, \pm 2)$, and $(\pm 3, \, 0)$.} so that the uniformly distributed FEC parity bits could be placed on the sign bits without changing $P_{|X_{\mathrm{c}}|}$.
The bit labelling for $128$-QAM was based on \cite[Fig.~3]{tenbrink_2004}.
For comparison, CCDM-based PS-$16$-QAM and PS-$64$-QAM were also evaluated, using the same PS overhead and codeword length as with compressed shaping.
We also evaluated the performance of PS-$4096$-QAM with an ideal Maxwell--Boltzmann input distribution and perfect data compression.

\begin{figure}[t]
	\begin{center}
		\setlength{\unitlength}{.6mm} %
		\scriptsize
		\vspace{-0.1cm}
		\includegraphics[scale=0.3]{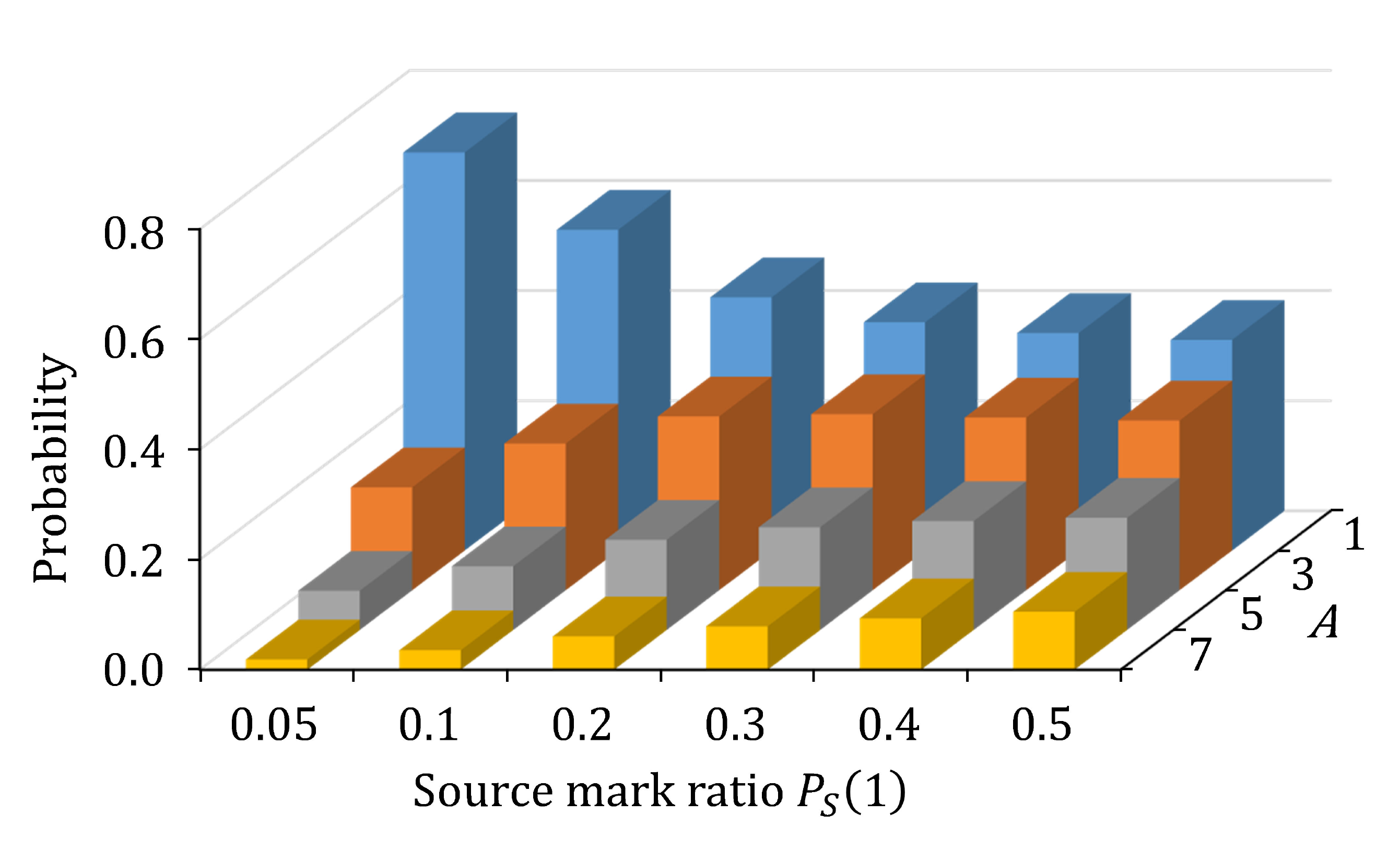} \\
		\vspace{-0.3cm}		
		\caption{An example of a one-dimensional amplitude PMF $P_{A}$ by compressed shaping $64$-QAM, for various source mark ratios $P_S(1)$. In situations with dynamically variable $P_S(1)$, the demapper assumes a constant $P_S(1) = 0.5$, as explained in the text.}
		\label{fig:pmf}
	\end{center}
	\vspace{-0.4cm}
\end{figure}

With CCDM, the energy $E_{\mathrm{2d}}$ in Fig.~\ref{fig:eavg} is constant for various $P_S(1)$ cases, because its output PMF $P_{A_{\mathrm{c}}}$ does not depend on incoming bits. With compressed shaping, on the other hand, $E_{\mathrm{2d}}$ decreases with decreasing $P_S(1)$ (and $\mathbb{H}(S)$) for each base constellation, although there are significant performance gaps to the ideal case (black solid line in Fig.~\ref{fig:eavg}), especially for high-order QAM.
\rev{This gap comes from the fixed compressed shaping regardless of the source statistics. While an adaptive compressed shaping would bring the performance closer to the ideal case, it would lead to a high complexity and is out of scope of this work.}
Fig.~\ref{fig:pmf} exemplifies the PMF $P_{A}$ for compressed shaping $64$-QAM with source mark ratios $P_S(1)=0.5$--$0.05$. Higher source nonuniformity (i.e., smaller $P_S(1)$) makes deeper PS. Under such source nonuniformity, we observed reduced $P_A(3)$, $P_A(5)$, and $P_A(7)$ and increased $P_A(1)$ compared with the uniform source case.
The transmitted PMFs of one-dimensional amplitudes $A$ and two-dimensional channel input symbols $X_{\mathrm{c}}$ at a given source mark ratio $P_S(1)$ are denoted as $P_{A(P_S(1))}$ and $P_{X_{\mathrm{c}}(P_S(1))}$, respectively.

\begin{figure}[t]
	\begin{center}
		\setlength{\unitlength}{.6mm} %
		\scriptsize
		\vspace{-0.1cm}
		\includegraphics[scale=0.32]{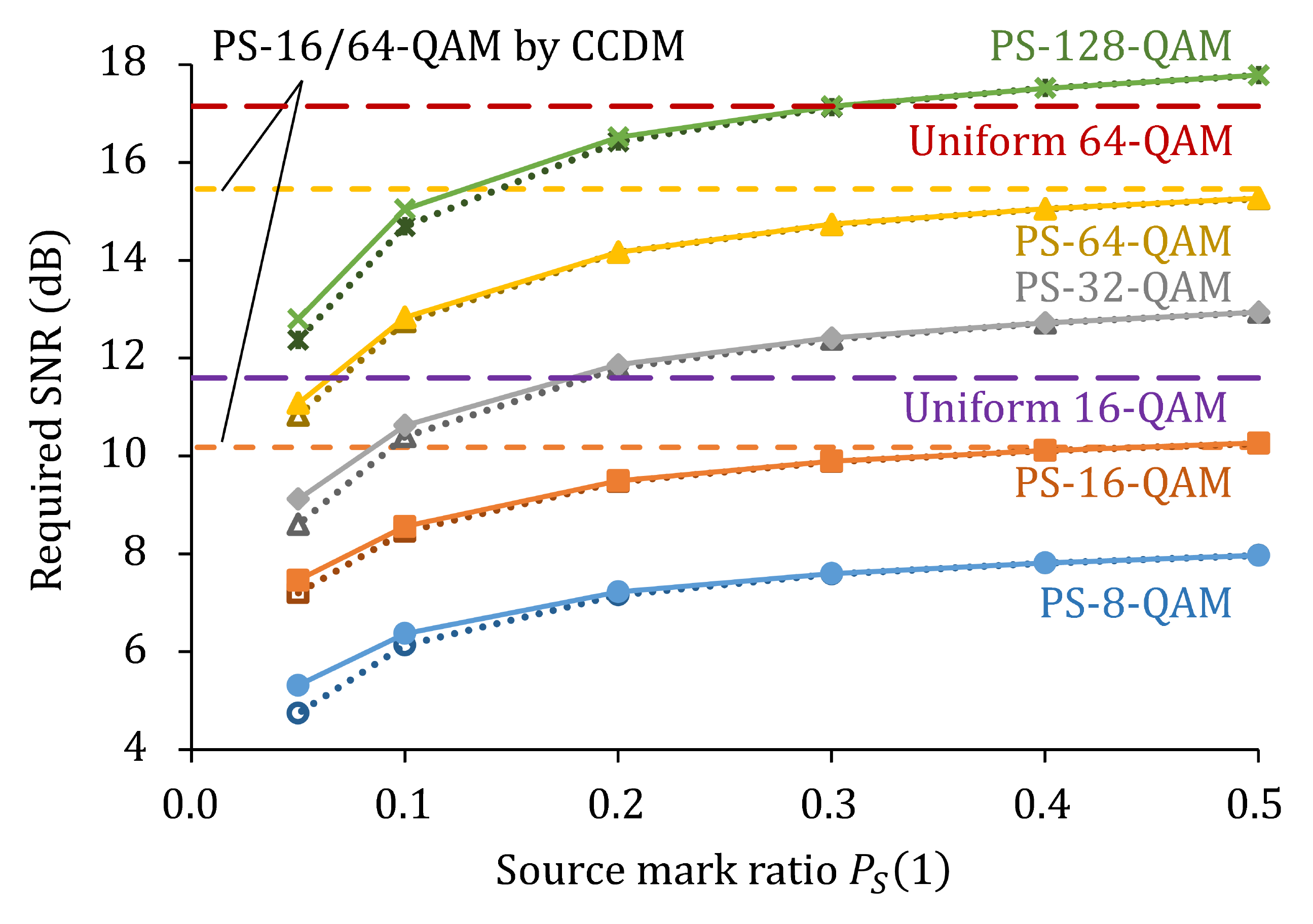} \\
		\vspace{-0.3cm}		
		\caption{Simulated required SNR for compressed shaping QAM under matched (dotted lines) and mismatched decoding (solid lines). In the mismatched case, we set the transmitted PMF assumed in the demapper to the one for a uniform source ($P_S(1)=0.5$), for every source mark ratio $P_S(1)$. Dashed and long-dashed lines correspond to CCDM and uniform QAM, respectively.}
		\label{fig:sim_rsnr}
	\end{center}
	\vspace{-0.4cm}
\end{figure}

We then simulated the required SNR over the Gaussian channel with the DVB-S2 low-density parity check code \cite{dvbs2} having a code rate of $5/6$, with a maximum number of decoding iterations of $50$.
Fig.~\ref{fig:sim_rsnr} shows the simulated required SNR with the maximum number of FEC decoding iterations. 
\rev{The required SNR is defined at an asymmetric information \cite{Yoshida:17,Yoshida:jlt20-2}\footnote{\rev{The asymmetric information is equivalent to normalized generalized mutual information \cite{Cho:17,cho_2019_jlt} and achievable FEC rate \cite{bocherer_2019_jlt} under matched decoding.}} of $0.87$, which corresponds to a pre-FEC BER of about $3.4\cdot 10^{-2}$. We observed no residual bit errors in more than $2.5 \cdot 10^6$ information bits after FEC decoding in the SNR regime larger than the required SNR for each case.}
We simulated both \emph{mismatched} and \emph{matched} decoding. The true transmitted two-dimensional symbol PMF is denoted by $P_{X_{\mathrm{c}}(P_S(1))}$ and the transmitted symbol PMF assumed in the soft demapping by $Q_{X_{\mathrm{c}}(P_S(1))}$. In the matched case, $Q_{X_{\mathrm{c}}(P_S(1))}$ was set to $P_{X_{\mathrm{c}}(P_S(1))}$ for all $P_S(1)$, while in the mismatched case, $Q_{X_{\mathrm{c}}(P_S(1))}$ was set to a fixed PMF of $P_{X_{\mathrm{c}}(0.5)}$ regardless of the true $P_S(1)$.
This is because under dynamically variable source situations, the true transmitted symbol PMF $P_{X_{\mathrm{c}}(P_S(1))}$ is hard to track in deployable systems.
As shown in Fig.~\ref{fig:sim_rsnr}, the required SNR can be reduced by compressed shaping, in contrast to the fixed required SNR by CCDM and uniform QAM with bit scrambling due to the fixed $P_{X_{\mathrm{c}}}$ ($=P_{X_{\mathrm{c}}(0.5)}$). The SNR penalty by the mismatch between $P_{X_{\mathrm{c}}(P_S(1))}$ and $Q_{X_{\mathrm{c}}(P_S(1))}$ is not significant except at very small $P_S(1)$.

\begin{figure}[t]
	\begin{center}
		\setlength{\unitlength}{.6mm} %
		\scriptsize
		\vspace{-0.1cm}
		\includegraphics[scale=0.32]{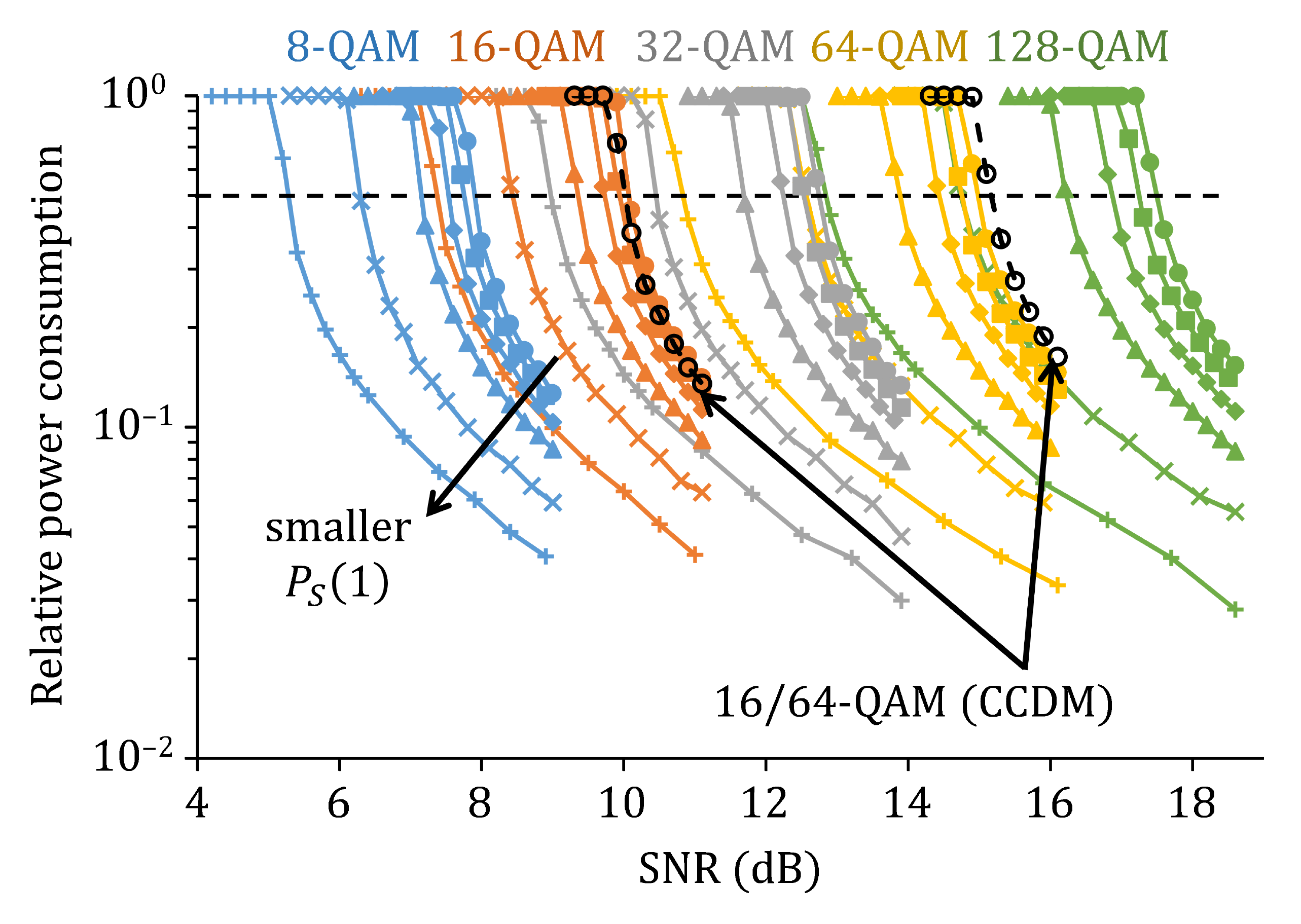} \\
		\vspace{-0.3cm}		
		\caption{Simulated relative power consumption of the FEC decoding, which is assumed to be proportional to average number of decoding iterations, as a function of SNR for PS-QAM. Solid and black dashed lines correspond to compressed shaping and CCDM, respectively.}
		\label{fig:sim_pcon}
	\end{center}
	\vspace{-0.4cm}
\end{figure}

The better performance of compressed shaping compared with conventional PS can be converted into lower power consumption.
We quantified the power consumption in FEC decoding, because it dominates the power consumption among all coding functions. 
Fig.~\ref{fig:sim_pcon} shows the relative power consumption of the FEC decoding for PS-QAM with $P_S(1) = 0.5$ (circle), $0.4$ (square), $0.3$ (diamond), $0.2$ (triangle), $0.1$ (cross), or $0.05$ (plus), so there are six curves for each QAM order with compressed shaping. The power consumption is assumed to be proportional to the average number of decoding iterations, which is almost proportional to the toggle rate in logical circuitry. The vertical axis in Fig.~\ref{fig:sim_pcon} is normalized by the maximum number of decoding iterations, i.e., $50$.
The soft demapping was assumed to be mismatched as in Fig.~\ref{fig:sim_rsnr}, i.e., $Q_{X_{\mathrm{c}}(P_S(1))}=P_{X_{\mathrm{c}}(0.5)}$. While CCDM consumes a fixed power even if $P_S(1)$ is reduced (this is the same for non-PS signaling with bit scrambling, but not shown in Fig.~\ref{fig:sim_pcon}), compressed shaping significantly reduces the power to about $10\%$ for highly nonuniform source probabilities, because of the smaller required SNR, which leads to a smaller number of decoding iterations than with conventional bit-scrambled PS.

\rev{While in this work, the source mark ratio is assumed to be constant, dynamic variations thereof were simulated in \cite[Sec.~5.4]{yoshida_2020_eic}. Source mark ratios were linearly varied, e.g., from $20$\% to $40$\% or from $10$\% to $50$\%, and compared with a constant $30$\% as a benchmark. 
For a larger range of variation, the performance improved slightly, even if the average source mark ratio over many blocks was the same.}
Note that additional complexity in compressed shaping is not significant when hierarchical DM is employed for shaping encoding/decoding, which will be shown in the next section.

\section{FPGA implementation}
\label{sec:fpga}

\begin{figure}[t]
	\begin{center}
		\setlength{\unitlength}{.6mm} %
		\scriptsize
		\vspace{-0.1cm}
		\includegraphics[scale=0.33]{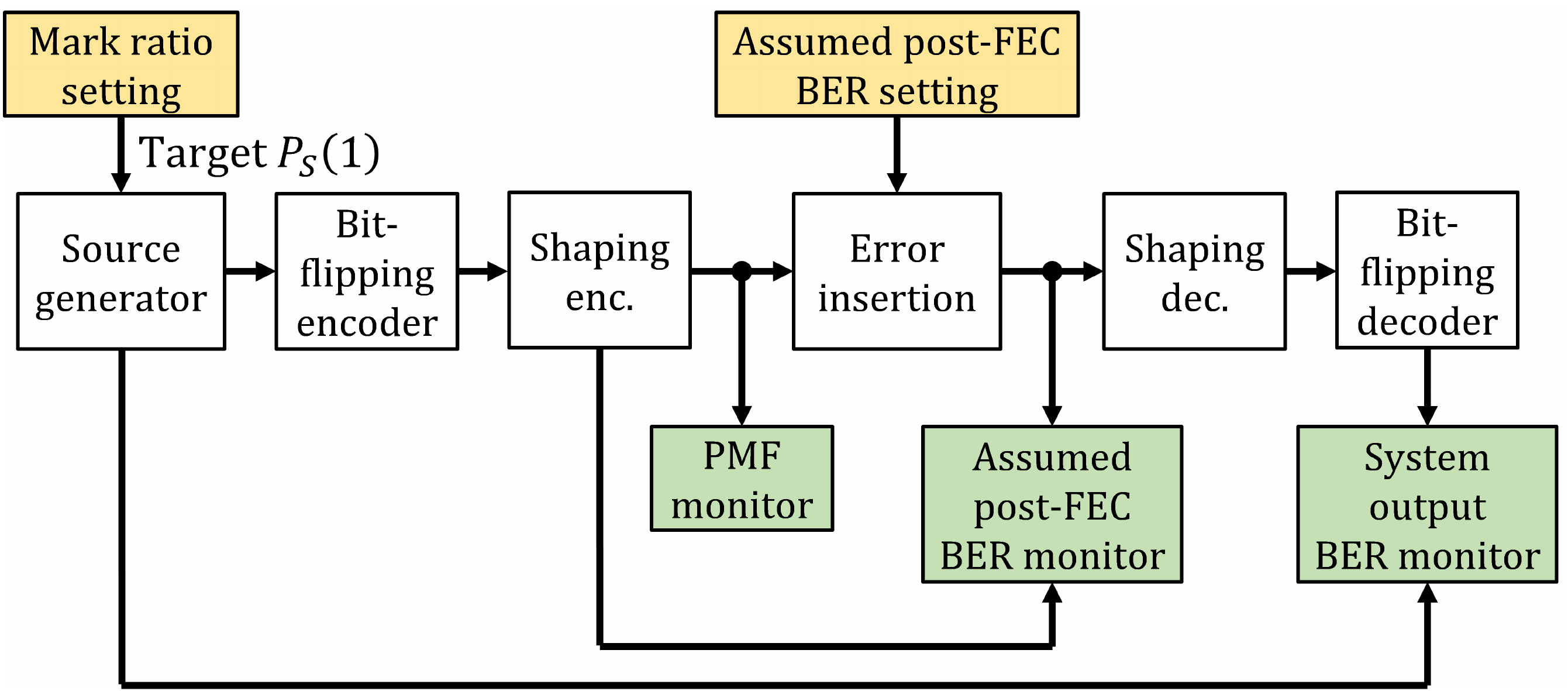} \\
		\vspace{-0.3cm}		
		\caption{Block diagram of FPGA implementation of compressed shaping.}
		\label{fig:fpga_imp}
	\end{center}
	\vspace{-0.4cm}
\end{figure}

\begin{figure}[t]
	\begin{center}
		\setlength{\unitlength}{.6mm} %
		\scriptsize
		\vspace{-0.1cm}
		\includegraphics[scale=0.2]{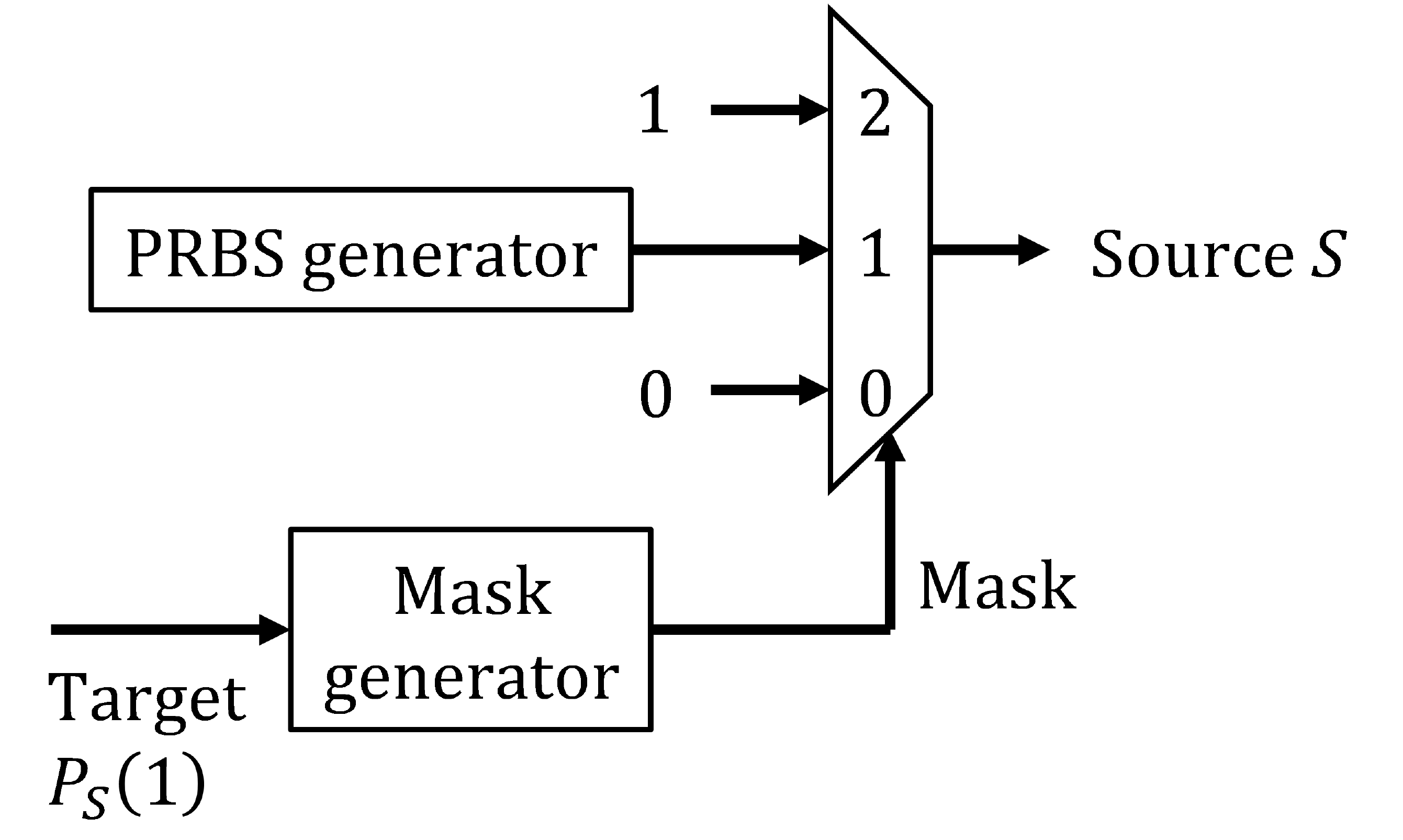} \\
		\vspace{-0.3cm}		
		\caption{Schematic of source generation function. The mask signal selects the output source bit from logic `$0$', `$1$', or a bit given by the PRBS based on the target source mark ratio $P_S(1)$.}
		\label{fig:sgen}
	\end{center}
	\begin{center}
		\setlength{\unitlength}{.6mm} %
		\scriptsize
		\vspace{-0.1cm}
		\includegraphics[scale=0.35]{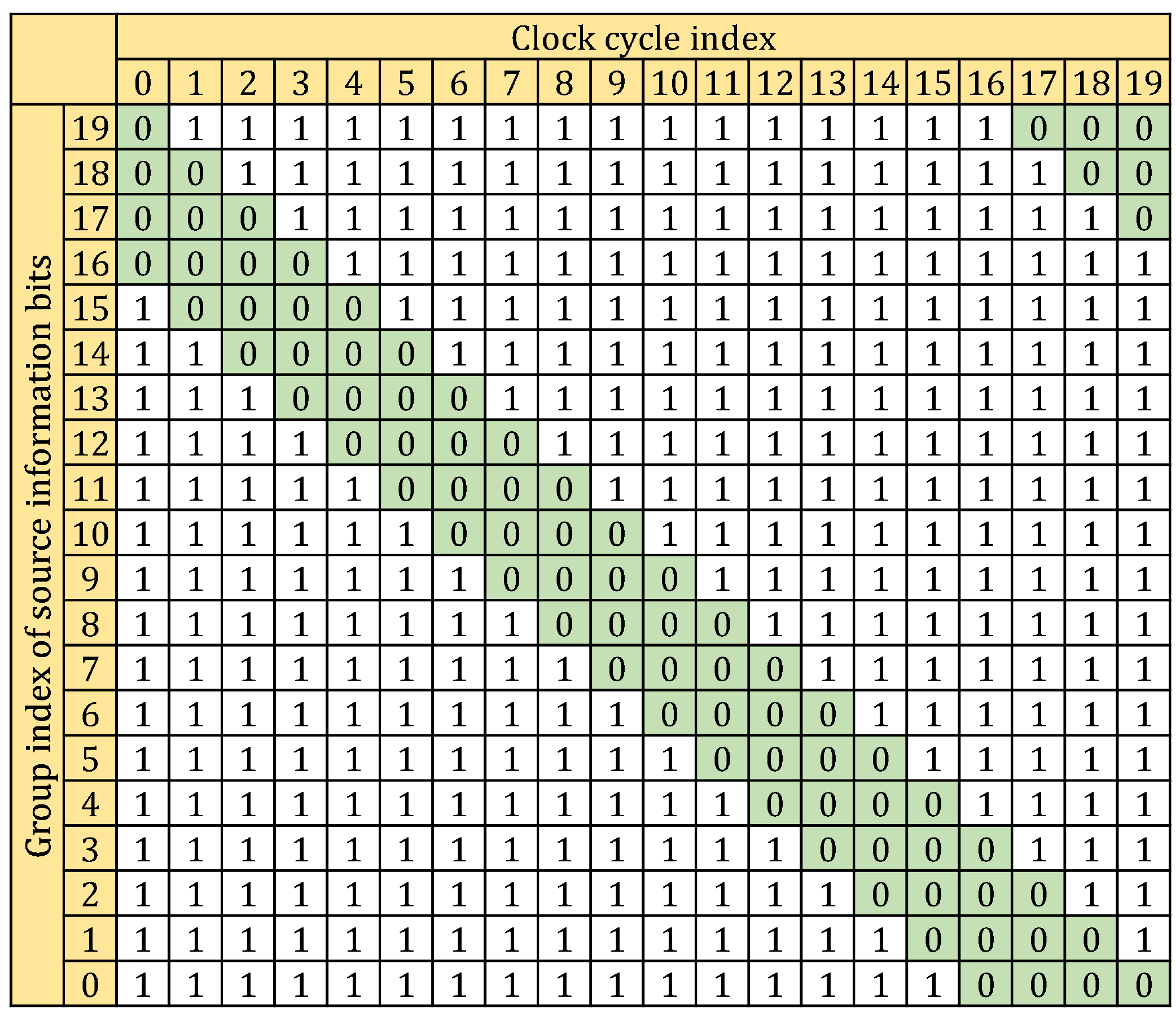} \\
		\vspace{-0.3cm}
		\caption{Example of mask signals for given clock cycle indices and group indices for a target source mark ratio $P_S(1)=0.4$. In each clock cycle index or each group index, $20\%$ of the generated bits are set directly to `$0$' and $80\%$ are taken from a uniform PRBS, which implies $20+80/2 = 60\%$ `$0$'s on average. }
		\label{fig:mask}
	\end{center}
	\vspace{-0.4cm}
\end{figure}

We implemented compressed shaping in a single FPGA chip on an evaluation board Xilinx\textregistered{} Virtex\textregistered{} Ultrascale+\texttrademark{} VCU118 XCVU9P. Fig.~\ref{fig:fpga_imp} shows the functional block diagram of the implemented circuitry. The source generator outputs source information bits based on a given target mark ratio $P_{S}(1)$. 
The schematic of the source generator is illustrated in Fig.~\ref{fig:sgen}.
Each source information bit is selected from one of three possible candidates; 0) the logic bit `$0$', 1) a PRBS of length $2^{31}-1$ bits, and 2) the logic bit `$1$'. The mask signal, used for the selection, is generated from the target source mark ratio $P_S(1)$, which is given by the user. If $P_S(1) \le 0.5$, the mask signal takes on values $0$ or $1$ such that the average fraction of logic `$1$'s in $S$ is $P_S(1)$, and if $P_S(1)>0.5$, the mask similarly is $1$ or $2$. 
Fig.~\ref{fig:mask} shows an exemplified mask signal when the target $P_S(1)$ is $0.4$. For simplicity, we classified source bits into 20 groups (32 bits per group), provided a `$0$' mask window for four groups, and slided the window in every clock cycle.

The bit-flipping encoder counts the numbers of `$0$'s and `$1$'s. When the number of `$1$'s is larger than that of `$0$'s, it flips all bits at the clock cycle and adds a parity bit `$1$'. Otherwise it just adds a parity bit `$0$'. The shaping encoding/decoding is realized by hierarchical DM having a total codeword length of $404$ bits for the shaped two-dimensional amplitudes. The number of shaped information bits per two-dimensional amplitude $k/n$ for compressed shaping employing hierarchical DM is generally flexible, and is in this implementation fixed to $372/202$ and $372/101$ for $16$-QAM and $64$-QAM, respectively, with compressed shaping.\footnote{A flexible choice of $(n,\,k)$ in an FPGA implementation of hierarchical DM is left as potential future work. The LUT contents can be reconfigured in software or firmware without increasing the RAM size.}
Before the receiver-side processing, we have an error insertion function, which inserts bit errors based on a given bit error rate (BER) before the shaping decoder. The shaping decoder is also implemented with a hierarchical DM decoder, and the bit-flipping decoder recovers the source bits. When the parity bit is `$1$', it flips all bits at the clock cycle and removes the parity bit. Otherwise, it just removes the parity bit. 

There are several monitoring functions, i.e., the PMF $P_{A_{\mathrm{c}}}$ at the shaping encoder output, the assumed post-FEC BER, and the system output BER. Note that hierarchical DM mainly consists of LUTs, which are implemented with random access memory (RAM). Because it is sensitive to unwanted bit-flipping due to radiation-induced soft errors, we implemented soft error protection circuitry.

The clock domain was initially single and the clock frequency $f_{\mathrm{clk}}$ for the fitting (FPGA synthesis) was $90\,\mathrm{MHz}$ \cite{yoshida_2019_ecoc_fpga}. Later we found a bottleneck in making the clock frequency higher inside the soft error protection circuitry (consisting of flip-flops and a selector tree for refreshing RAM contents), so we separated the clock domain into one for data processing and one for control. We then achieved $f_{\mathrm{clk}}=240\,\mathrm{MHz}$ for the data processing for higher throughput. 
Assuming a suitable FEC concatenation (not implemented here), the system throughputs for compressed shaping 16- and $64$-QAM would be $57$ and $42\,\mathrm{Gb/s}$ at $f_{\mathrm{clk}} = 90\,\mathrm{MHz}$, and $153$ and $113\,\mathrm{Gb/s}$ at $f_{\mathrm{clk}} = 240\,\mathrm{MHz}$, respectively. The number of bits per PS codeword were the same in both cases, i.e., there were half as many PS-$64$-QAM symbols as PS-$16$-QAM symbols.

\begin{figure}[t]
	\begin{center}
		\setlength{\unitlength}{.6mm} %
		\scriptsize
		\vspace{-0.1cm}
		\includegraphics[scale=0.32]{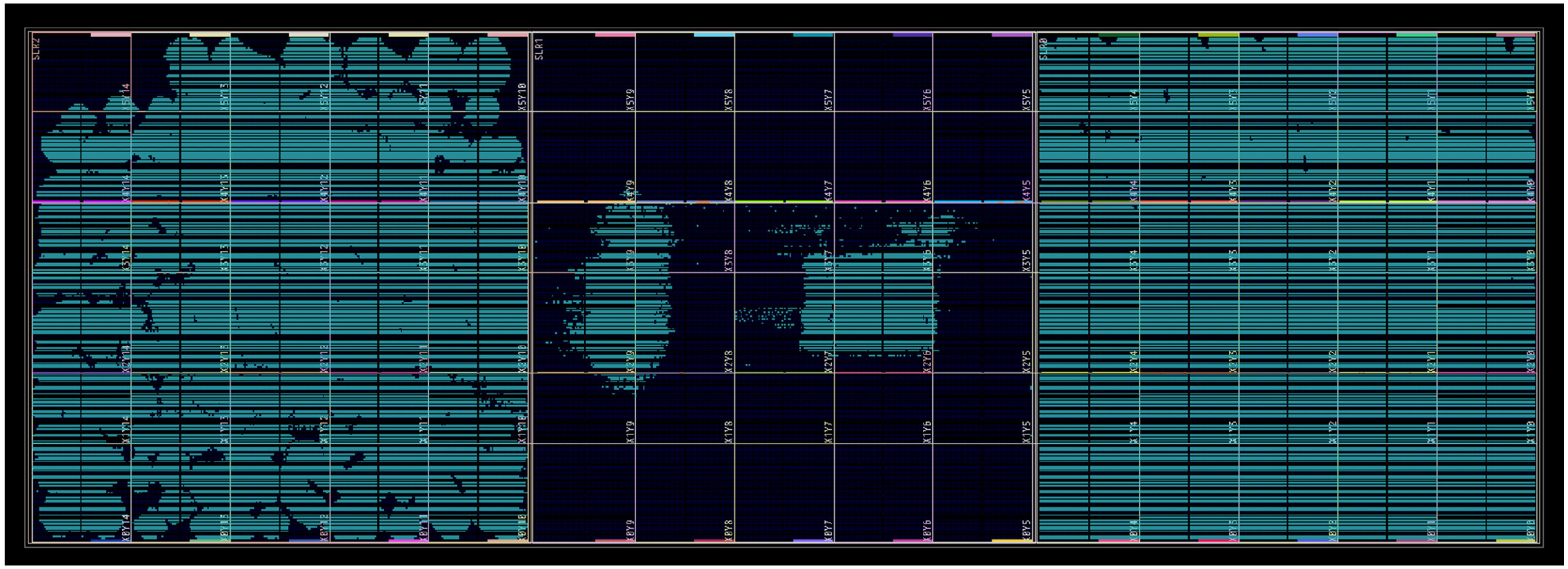} \\
		\vspace{-0.3cm}		
		\caption{Utilized area (green) of a single FPGA chip for compressed shaping.}
		\label{fig:util_area}
	\end{center}
	\vspace{-0.4cm}
\end{figure}

Tab.~\ref{tab:util} shows the utilized hardware resources at a clock frequency of the data signals $f_{\mathrm{clk}}=90$ or $240\,\mathrm{MHz}$. Fig.~\ref{fig:util_area} depicts the utilized area of the FPGA chip having three dies. The used resource elements were mainly located in the left and right dies, and the center die was mainly used for connection between the two dies. The \emph{register} elements were mainly used by the soft error protection circuitry for storing the RAM contents.
Out of the about 290,000 utilized \emph{LUT as logic} elements, $70\%$ were provided to external functions such as source generator and BER/PMF monitors, $10\%$ were for bit-flipping encoding, and the rest was for other combinational logics.
The data processing of the hierarchical DM used totally $3.5\,\mathrm{Mb}$ \emph{block RAM} elements. No \emph{ultra RAM} or \emph{DSP slice} elements were used.

As a benchmark, 400 ZR FEC was implemented in $50$  Xilinx\textregistered{} Virtex\textregistered{} Ultrascale\texttrademark{} FPGAs \cite{cai_2018_ofc}. The clock frequency was $125\,\mathrm{MHz}$ and the system throughput was $200\,\mathrm{Gb/s}$ with external functions, i.e., source generation, mapping, demapping, interleaver, de-interleaver, and noise loading. Even considering that this reference includes many external functions, compressed shaping utilizes a very small amount of hardware resources.

Tab.~\ref{tab:pow} shows the estimated dynamic power consumption for compressed shaping $64$-QAM at $f_{\mathrm{clk}}=240\,\mathrm{MHz}$. While we estimated a static power consumption based on a default toggle rate of $12.5\%$ in our previous report \cite{yoshida_2019_ecoc_fpga}, we now improved the estimation accuracy by taking realistic node switching activities into account based on register transfer level simulation waveform (a so-called switching activity interchange format) over $4000$ clock cycles, where the source mark ratio $P_S(1)$ was set to $0.5$. 

The blocks of Tx data and Rx data in Tab.~\ref{tab:pow} are essential for data communications.
Among them, the preprocessing (including bit-flipping encoding) and DM encoder core in the Tx data block, and the DM decoder core in the Rx data block, consumed most of the power. The DM encoder and decoder cores mainly consisted of block RAMs for LUTs. The bit-flipping encoding counted the logic `$1$'s using adders, leading to a relatively large power consumption. Other pre- and postprocessing functions including the lane reorder consumed little power. Note that the power consumption of either the Tx or the Rx data block was smaller than that of external functions, e.g., source generation and BER monitoring.
We had nonnegligible power consumption in the control blocks. They had soft error protection functions for the DM encoder and decoder cores by holding copies of the entire RAM contents in registers and refreshing the RAM intermittently. These powers can be expected to be less in an ASIC implementation by making the activation ratio small.

\begin{table}[t]
	\caption{Utilization of key resources in FPGA for compressed shaping at $f_{\mathrm{clk}} = 90$ or $240\,\mathrm{MHz}$.}
	\label{tab:util}
	\vspace{-0.4cm}
	\begin{center}
	\begin{tabular}{llrrr}
		\hline
		\multicolumn{1}{l}{Category} & \multicolumn{1}{l}{Element} & \multicolumn{1}{c}{Available} & \multicolumn{2}{c}{Utilization} \\
		& & & $90\,\mathrm{MHz}$ & $240\,\mathrm{MHz}$ \\
		\hline\hline
		System logic & LUT as logic & $1182\mathrm{k}$ & $24.17\%$ & $24.64\%$ \\ 
		cell & Register & $2364\mathrm{k}$ & $22.49\%$ & $21.38\%$ \\ \hline
		Memory & Block RAM & $75.9\,\mathrm{Mb}$ & $4.65\%$ & $4.65\%$ \\
		& Ultra RAM & $270.0\,\mathrm{Mb}$ & Not used & Not used \\ \hline
		DSP slice &  & $6840\,\mathrm{slices}$ & Not used & Not used \\ \hline
	\end{tabular}
	\end{center}
\end{table}

\begin{table}[t]
	\caption{Simulated dynamic power consumption of compressed shaping $64$-QAM at $f_{\mathrm{clk}} = 240\,\mathrm{MHz}$.}
	\label{tab:pow}
	\vspace{-0.4cm}
	\begin{center}
	\begin{tabular}{llr}
		\hline
		Block & Function & Power (mW) \\
		\hline\hline
		External & Source generation & $589$ \\
		functions & Error insertion & $11$ \\
		& BER monitor & $638$ \\
		& Clock generation & $99$ \\
		& Sub-total & $1\,337$ \\ \hline
		Tx data & Pre-processing & $232$ \\
		& DM encoder core & $456$ \\
		& Post-processing & $78$ \\
		& Delay adjustment & $46$ \\
		& Sub-total & $812$ \\ \hline
		Rx data & Pre-processing & $44$ \\
		& DM decoder core & $562$ \\
		& Post-processing & $34$ \\
		& Delay adjustment & $35$ \\
		& Sub-total & $675$ \\ \hline
		Tx control & & $320$ \\ \hline
		Rx control & & $295$ \\ \hline
		Tx/Rx control & & $167$ \\ \hline
		Total & & $3\,613$ \\ \hline
	\end{tabular}
	\end{center}
\end{table}

\section{\sout{Realtime}\rev{Hardware} demonstration}
\label{sec:demo}

\begin{figure}[t]
	\begin{center}
		\setlength{\unitlength}{.6mm} %
		\scriptsize
		\vspace{-0.1cm}
		\includegraphics[scale=0.32]{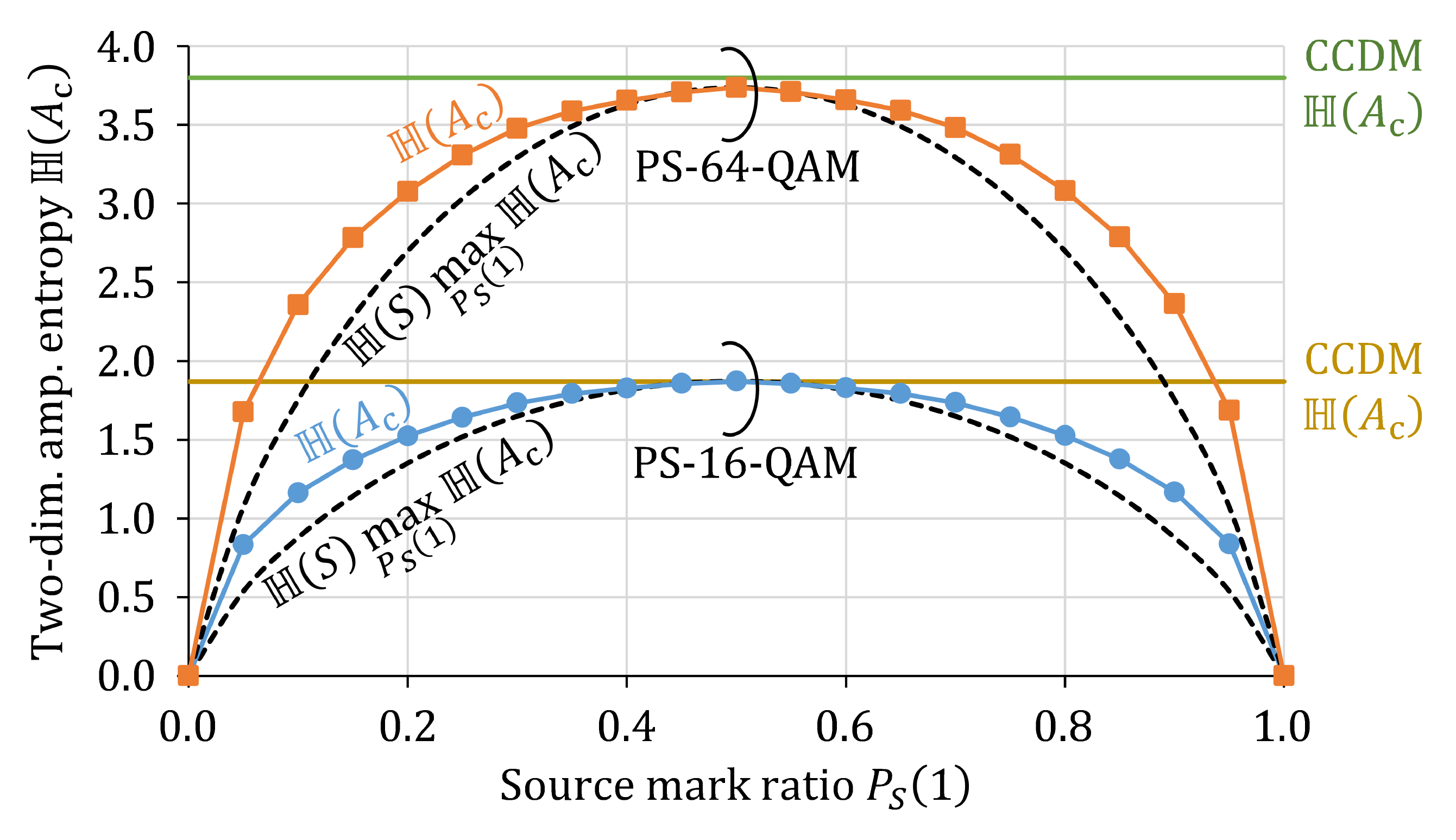} \\
		\vspace{-0.3cm}		
		\caption{Measured two-dimensional amplitude entropy $\mathbb{H}(A_{\mathrm{c}})$ for compressed shaping as a function of the target source mark ratio $P_S(1)$ based on $>10^{10}$ two-dimensional amplitude samples. The dashed lines show the binary entropy function scaled to the maximum entropies of the two-dimensional amplitudes for comparison.}
		\label{fig:syment}
	\end{center}
	\vspace{-0.4cm}
\end{figure}

We \sout{made real-time evaluations of} \rev{demonstrated} compressed shaping by employing the FPGA evaluation board at $f_{\mathrm{clk}} = 90\,\mathrm{MHz}$. 
First, the transmitter-side functions were verified. Histograms of the two-dimensional amplitude $A_{\mathrm{c}}$ for compressed shaping $16$- and $64$-QAM were measured over $10^{10}$ two-dimensional amplitude samples for various source mark ratios $P_S(1)$. The obtained histograms were interpreted as PMFs $P_{A_{\mathrm{c}}}$ and their entropies were computed. Fig.~\ref{fig:syment} shows $\mathbb{H}(A_{\mathrm{c}})$ as a function of $P_S(1)$. The entropy $\mathbb{H}(A_{\mathrm{c}})$ is maximum for $P_S(1)=0.5$. The two-dimensional PS rate losses $R_{\mathrm{loss}}(U,A_{\mathrm{c}})$ are $0.034$ and $0.064\,\mathrm{bpcu}$ for the exemplified PS-$16$-QAM and PS-$64$-QAM schemes, respectively.
When $P_S(1)$ deviates from $0.5$, i.e., $\mathbb{H}(S)$ decreases, $\mathbb{H}(A_{\mathrm{c}})$ also decreases monotonically. Because of the bit-flipping encoding, $\mathbb{H}(A_{\mathrm{c}})$ is symmetric around $P_S(1)=0.5$. In contrast, the CCDM-based schemes show a constant $\mathbb{H}(A_{\mathrm{c}})$, which is independent of the source distribution.
Scaled binary entropy functions $\mathbb{H}(S) \max_{P_S(1)}\mathbb{H}(A_{\mathrm{c}})$ are also depicted in Fig.~\ref{fig:syment} with black dotted lines for the two cases. The gap between $\mathbb{H}(A_{\mathrm{c}})$ and $\mathbb{H}(S) \max_{P_S(1)}\mathbb{H}(A_{\mathrm{c}})$ is a data compression rate loss. 
The data compression rate loss is ideally zero but is larger here for smaller $\mathbb{H}(S)$, due to the nonideal simple processing of compressed shaping. Here the bit sequence conversion loss in compressed shaping $R_{\mathrm{loss}}(S_{\mathrm{a}},A_{\mathrm{c}})$ in \eqref{eq:rloss_cps} is given by the sum of rate losses in PS and data compression. Regardless of the rate loss increase in data compression, $\mathbb{H}(A_{\mathrm{c}})$ itself is reduced to help reduce the required SNR substantially.

\begin{figure}[t]
	\begin{center}
		\setlength{\unitlength}{.6mm} %
		\scriptsize
		\vspace{-0.1cm}
		\includegraphics[scale=0.32]{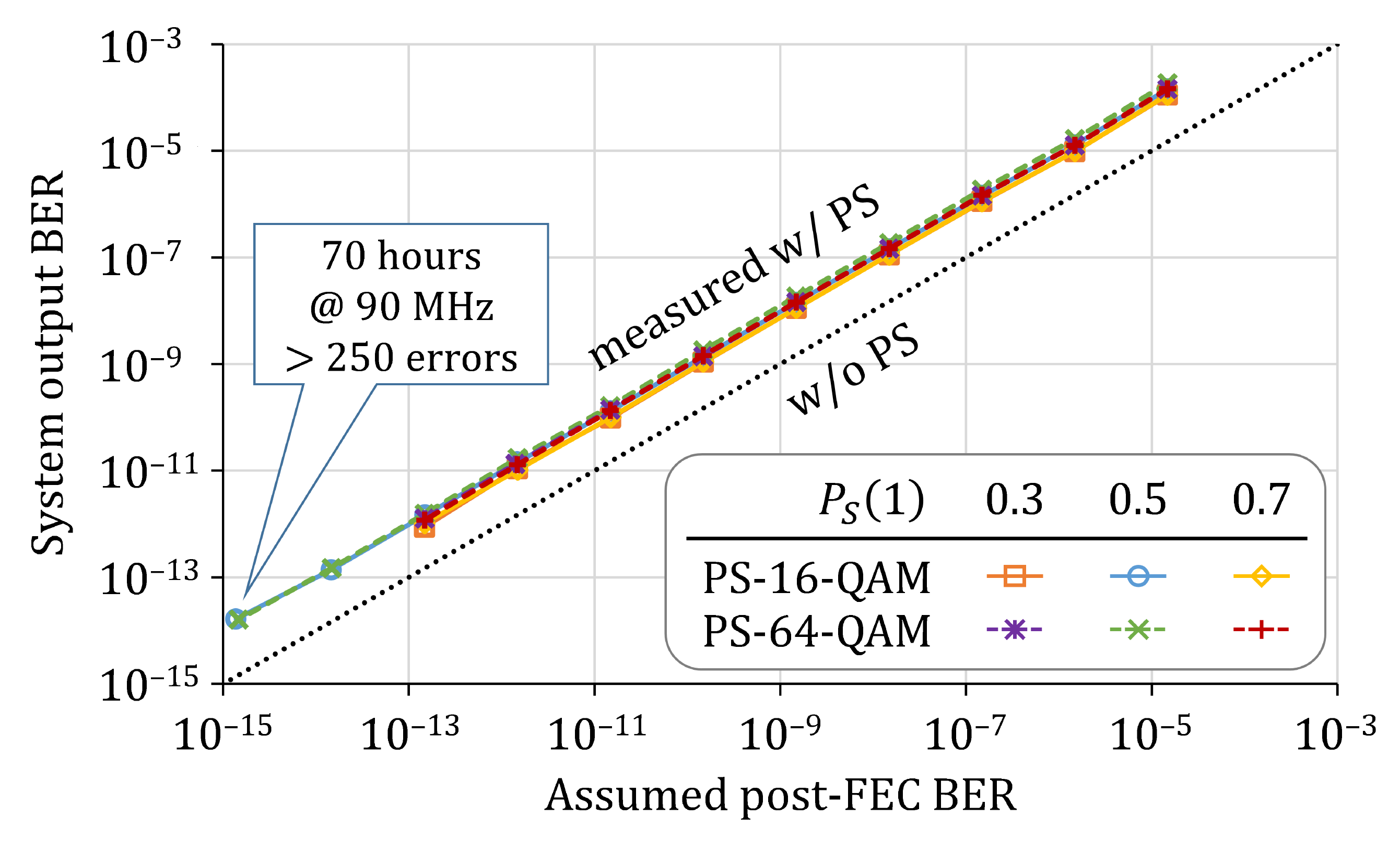} \\
		\vspace{-0.3cm}		
		\caption{Measured system output BER for compressed shaping as a function of assumed post-FEC BER in back-to-back error insertion tests.}
		\label{fig:ber}
	\end{center}
	\vspace{-0.4cm}
\end{figure}

Next, the receiver-side functions were verified. We turned the error insertion function on to provide sparse bit errors between the transmitter and the receiver. As reported in \cite{yoshida_2019_jlt}, sparse errors creates the worst system output BER at a given post-FEC BER. We swept the assumed post-FEC BER from $10^{-5}$ to $10^{-15}$ and examined $P_S(1)$ of $0.3$, $0.5$, and $0.7$ for compressed shaping $16$- and $64$-QAM. After the receiver-side processing, the system output BER was measured.
Fig.~\ref{fig:ber} shows the system output BER as a function of assumed post-FEC BER in the back-to-back error insertion test. The BER increase due to compressed shaping decoding was only around $10$ times because hierarchical DM can partially decode correctly even if there are incoming bit errors. As predicted in \cite{yoshida_2019_jlt}, the BER increase factor is significantly smaller than for other DM techniques. For example, CCDM decoding having a PS codeword length of around $400$ bits results in more than $100$ times higher BER, though we could not implement CCDM in the FPGA due to its high complexity.
We performed $70$ hours of long-term measurement only in the case of $P_S(1)=0.5$, due to time constraints. The number of observed bit errors after compressed shaping decoding were more than $250$. The system output BER was $1.5\cdot 10^{-14}$ at a post-FEC BER of $1.6\cdot 10^{-15}$.
This proves that the proposed compressed shaping does not cause an error floor or excessive error increase, so the required post-FEC BER remains around $10^{-16}$ to satisfy a required system output BER of $10^{-15}$.

\section{Summary}
\label{sec:sum}
We proposed and demonstrated compressed shaping, i.e., the application of hierarchical DM to simultaneous source data compression and probabilistic shaping, which is an example of joint source--channel coding. Under a reduced source entropy, compressed shaping reduces channel input symbol entropy, symbol energy, and required SNR. Simulation results showed its smaller required SNR, as well as reduced power consumption compared with CCDM. We implemented compressed shaping, which are mainly hierarchical DM encoding/decoding, into a single FPGA and estimated the dynamic power consumption based on simulated waveforms. The system throughput reached $153$ and $113\,\mathrm{Gb/s}$ for compressed shaping $16$- and $64$-QAM, respectively. Real-time evaluation results showed expected performance in both encoding and decoding. Compressed shaping works at a very small BER of around $10^{-15}$ without any error floor, and its decoding increases the BER only around $10$ times, which is small compared to other DM schemes.

\rev{The time-varying symbol energy in compressed shaping causes practical issues with respect to electrical amplitude, optical power, and SNR control in fiber-optic communication systems, although such a variation per channel can be statistically relaxed by multiplexing many channels. 
One example can be found in a wavelength-division-multiplexed system operated at a given total launch power.
Without any control of the average symbol energy per channel, the average optical power in some channels can temporarily be reduced when the source entropy becomes small. This leads to a larger average optical power and thus higher nonlinearity impairments in other copropagating channels. 
Analyses of such issues and development of appropriate control methods are deferred to a future work.
We can also potentially explore the adaption of compressed shaping to mass trends of source and channel variations.}

\section*{Acknowledgment}
We thank Kyo Inoue of Osaka University for assistance in the research.

\balance


\end{document}